\documentclass[10pt,oneside]{article}

\usepackage[english]{babel}

\usepackage{amsmath,amssymb,amsfonts}
\usepackage[utf8]{inputenc}
\usepackage[T1]{fontenc}
\usepackage{stix}
\usepackage[scaled]{helvet}
\usepackage[scaled]{inconsolata}

\usepackage{lastpage}

\usepackage{setspace}

\usepackage{ccicons}

\usepackage[hang,flushmargin]{footmisc}

\usepackage{geometry}

\usepackage{fancyhdr}

\makeatletter
\newcounter{tableno}
\newenvironment{tablenos:no-prefix-table-caption}{
  \caption@ifcompatibility{}{
    \let\oldthetable\thetable
    \let\oldtheHtable\theHtable
    \renewcommand{\thetable}{tableno:\thetableno}
    \renewcommand{\theHtable}{tableno:\thetableno}
    \stepcounter{tableno}
    \captionsetup{labelformat=empty}
  }
}{
  \caption@ifcompatibility{}{
    \captionsetup{labelformat=default}
    \let\thetable\oldthetable
    \let\theHtable\oldtheHtable
    \addtocounter{table}{-1}
  }
}
\makeatother

\usepackage{array}
\newcommand{\PreserveBackslash}[1]{\let\temp=\\#1\let\\=\temp}

\usepackage[breaklinks=true]{hyperref}
\hypersetup{colorlinks,%
citecolor=blue,%
filecolor=blue,%
linkcolor=blue,%
urlcolor=blue}
\usepackage{url}

\usepackage{caption}
\usepackage{cleveref}

\usepackage{graphicx}
\makeatletter
\def\maxwidth{\ifdim\Gin@nat@width>\linewidth\linewidth
\else\Gin@nat@width\fi}
\makeatother
\let\Oldincludegraphics\includegraphics
\renewcommand{\includegraphics}[1]{\Oldincludegraphics[width=\maxwidth]{#1}}

\usepackage{longtable}
\usepackage{booktabs}

\usepackage{color}
\usepackage{fancyvrb}

\DefineVerbatimEnvironment{Highlighting}{Verbatim}{commandchars=\\\{\}}
\usepackage{framed}
\definecolor{shadecolor}{RGB}{248,248,248}

\newlength{\cslhangindent}
\newlength{\csllabelwidth}
\newenvironment{CSLReferences}[3] 
 {
  \setlength{\parindent}{0pt}
  \ifodd #1 \everypar{\setlength{\hangindent}{\cslhangindent}}\ignorespaces\fi
  \ifnum #2 > 0
  \setlength{\parskip}{#2\baselineskip}
  \fi
 }%
 {}
\usepackage{calc} 

\usepackage[table,dvipsnames]{xcolor}

\usepackage[singlelinecheck=off]{caption}
\usepackage{floatrow}
\usepackage{titlesec}

\titleformat{\section}[block]
{\normalfont\large\sffamily}
{\thesection}{.5em}{\titlerule\\[.8ex]\bfseries}

\titleformat{\subsection}[runin]
{\normalfont\fontseries{b}\selectfont\filright\sffamily}
{\thesubsection.}{.5em}{}

\titleformat{\subsubsection}[runin]
{\normalfont\itshape\rmfamily\bfseries}{\thesubsubsection}{1em}{}

\fancypagestyle{firstpage}
{
   \fancyhf{}
   
   \fancyfoot[R]{\footnotesize\ccby}
   \fancyfoot[L]{\footnotesize\sffamily\today}
}

\fancypagestyle{normal}
{
  \fancyhf{}
  \fancyfoot[R]{\footnotesize\sffamily\thepage\ of \pageref*{LastPage}}
}

\usepackage{tikz}
\begin{document}
\tikz [remember picture, overlay] %
\node [shift={(-0.6in,1.1cm)},scale=0.2,opacity=0.4] at (current page.south east)[anchor=south east]{\includegraphics{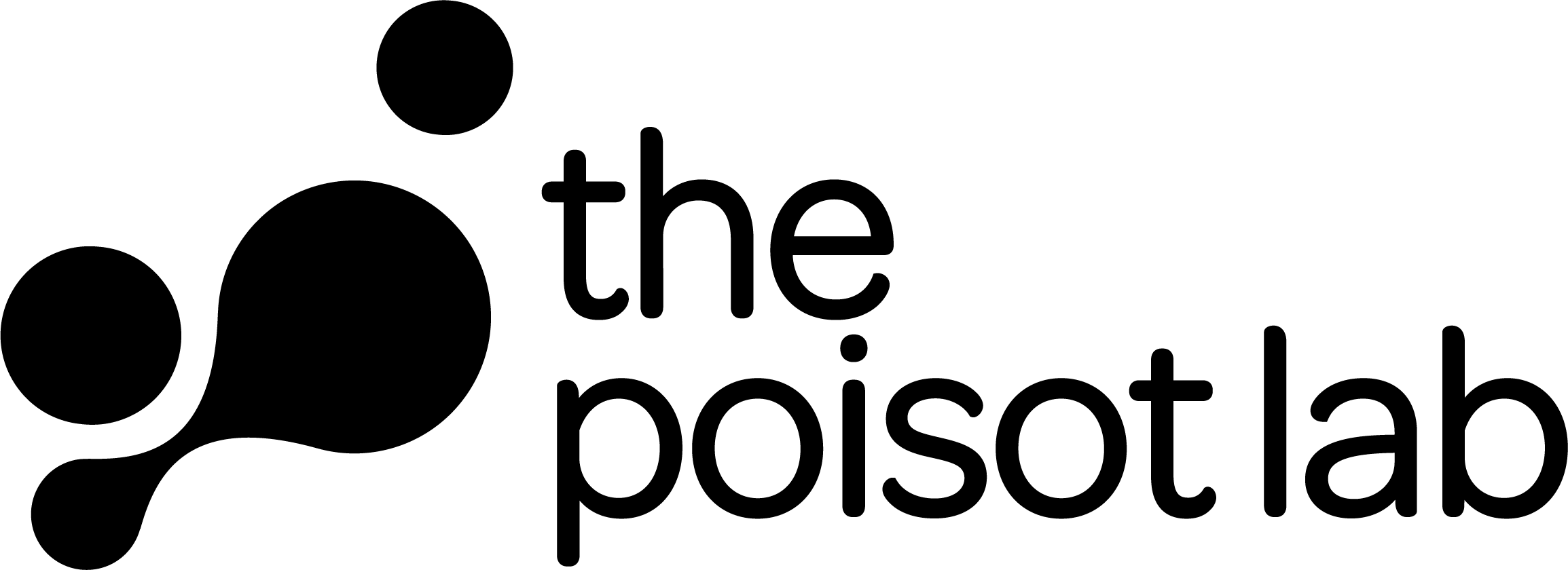}};%
\pagestyle{normal}
\thispagestyle{firstpage}

\newcommand{\colorRule}[3][black]{\textcolor[HTML]{#1}{\rule{#2}{#3}}}

\noindent {\LARGE \textbf{\textsf{What constrains food webs? A maximum
entropy framework for predicting their structure with minimal biases}}}

\medskip
\begin{flushleft}
{\small
\href{https://orcid.org/0000-0001-9051-0597}{Francis\,Banville}%
\,\textsuperscript{1,2,3}, %
\href{https://orcid.org/0000-0002-4498-7076}{Dominique\,Gravel}%
\,\textsuperscript{2,3}, %
\href{https://orcid.org/0000-0002-0735-5184}{Timothée\,Poisot}%
\,\textsuperscript{1,3}
\vskip 1em
\textsuperscript{1}\,Université de
Montréal; \textsuperscript{2}\,Université de
Sherbrooke; \textsuperscript{3}\,Quebec Centre for Biodiversity
Science\\
\vskip 1em
\textbf{Correspondance to:}\\
Francis Banville --- \texttt{francis.banville@umontreal.ca}\\
}
\end{flushleft}

\vskip 2em
\makebox[0pt][l]{\colorRule[CCCCCC]{2.0\textwidth}{0.5pt}}
\vskip 2em
\noindent

\marginpar{\vskip 1em\flushright
{\small{\bfseries Keywords}:\par
ecological modelling\\ecological networks\\food webs\\maximum
entropy\\null models\\}
}

\textbf{Abstract}:\,Food webs are complex ecological networks whose
structure is both ecologically and statistically constrained, with many
network properties being correlated with each other. Despite the
recognition of these invariable relationships in food webs, the use of
the principle of maximum entropy (MaxEnt) in network ecology is still
rare. This is surprising considering that MaxEnt is a statistical tool
precisely designed for understanding and predicting many different types
of constrained systems. Precisely, this principle asserts that the
least-biased probability distribution of a system's property,
constrained by prior knowledge about that system, is the one with
maximum information entropy. MaxEnt has been proven useful in many
ecological modelling problems, but its application in food webs and
other ecological networks is limited. Here we show how MaxEnt can be
used to derive many food-web properties both analytically and
heuristically. First, we show how the joint degree distribution (the
joint probability distribution of the numbers of prey and predators for
each species in the network) can be derived analytically using the
number of species and the number of interactions in food webs. Second,
we present a heuristic and flexible approach of finding a network's
adjacency matrix (the network's representation in matrix format) based
on simulated annealing and SVD entropy. We built two heuristic models
using the connectance and the joint degree sequence as statistical
constraints, respectively. We compared both models' predictions against
corresponding null and neutral models commonly used in network ecology
using open access data of terrestrial and aquatic food webs sampled
globally (N = 257). We found that the heuristic model constrained by the
joint degree sequence was a good predictor of many measures of food-web
structure, especially the nestedness and motifs distribution.
Specifically, our results suggest that the structure of terrestrial and
aquatic food webs is mainly driven by their joint degree distribution.

\vskip 2em
\makebox[0pt][l]{\colorRule[CCCCCC]{2.0\textwidth}{0.5pt}}
\vskip 2em

\hypertarget{introduction}{%
\section{Introduction}\label{introduction}}

\hypertarget{the-constrained-structure-of-ecological-networks}{%
\subsection{The constrained structure of ecological
networks}\label{the-constrained-structure-of-ecological-networks}}

A variety of measures of the structure of ecological networks have been
used to describe the organization of species interactions in a
biological community (Delmas et al. 2019). These measures provide
valuable information on the functioning of ecosystems and their
responses to environmental change (e.g., Pascual and Dunne 2006; Gómez,
Perfectti, and Jordano 2011). For instance, Bascompte et al. (2003)
suggest that plant--pollinator and seed-disperser networks have a highly
nested structure that can promote species persistence. Another example,
in food webs, shows that a high connectance can promote the robustness
of the system to species lost (Dunne, Williams, and Martinez 2002).
However, despite the growing literature on the ecological implications
of network structure, the association between many of these measures
impedes our ability to fully understand what drives the structure and
behavior of ecological networks. In particular, nestedness and
modularity are strongly associated in ecological networks (Fortuna et
al. 2010), and network connectance has been shown to be an important
driver of many other emerging network properties (Poisot and Gravel
2014). In light of these observations, it is difficult to assess whether
attributed effects of given properties are the artifacts of other,
perhaps simpler, measures.

One way to tackle this issue is first to recognize that food webs and
other ecological networks are constrained systems. In other words, the
space of possible network configurations shrinks as we know more about a
network structure. For example, there is a finite number of networks
with specified numbers of nodes and edges. Indeed, the structure of
ecological networks is first and above all constrained by the number of
species, or nodes, present. Food webs with high species richness
typically have a lower connectance (MacDonald, Banville, and Poisot
2020) than smaller networks. This is because the number of realized
interactions in empirical food webs scales slower than the number of
possible species pairs (MacDonald, Banville, and Poisot 2020). As shown
by Poisot and Gravel (2014), connectance itself can constrain different
aspects of network structure such as the degree distribution (i.e.~the
probability distribution of the number of interspecific interactions
realized by a species). Other measures, such as the maximum trophic
level, can also constrain the space of feasible networks.

Prior knowledge on the structure of ecological networks is thus
especially useful in the current context of data scarcity about species
interactions (Jordano 2016). Indeed, this Eltonian shortfall (Hortal et
al. 2015) can be partially alleviated using known information about an
ecological network. As suggested by Strydom et al. (2021), network
structure can be used to improve the prediction of pairwise species
interactions when data is lacking by constraining the space of feasible
networks. Similarly, partial knowledge on the structure of an ecological
network can also be used to predict others of its properties by
constraining their range of possible values. This is important given
that many aspects of network structure cannot be measured empirically
without data on pairwise species interactions, a prevailing situation
worldwide (Poisot et al. 2021).

Understanding the ecological constraints that shape species interactions
networks and predicting their emerging structure are thus two
complementary aims of network ecology. This distinction between
understanding and predicting is essential when using statistical and
mathematical models in network ecology and interpreting them. On one
hand, null models help us identify potential ecological mechanisms that
drive species interactions and constrain ecological networks. Null
models generate a distribution for a target measure using a set of rules
that exclude the mechanism of interest (Fortuna and Bascompte 2006;
Delmas et al. 2019). The deviation between the model and empirical data
helps us evaluate the effect of this ecological process in nature. On
the other hand, predictive models can help fill many gaps on species
interactions data. A variety of such models have recently been developed
using machine learning and other statistical tools, most of which are
presented in Strydom et al. (2021). However, given the constrained
nature of ecological networks, it is surprising that the principle of
maximum entropy, a mathematical method designed for both the analysis
and prediction of constrained systems, has been barely used in network
ecology.

\hypertarget{the-principle-of-maximum-entropy-a-primer-for-ecologists}{%
\subsection{The principle of maximum entropy: A primer for
ecologists}\label{the-principle-of-maximum-entropy-a-primer-for-ecologists}}

The principle of maximum entropy (MaxEnt) is a mathematical method of
finding probability distributions, strongly rooted in statistical
mechanics and information theory (Jaynes 1957a, 1957b; Harremoës and
Topsøe 2001). Starting from a set of constraints given by prior
knowledge of a system (i.e.~what we call state variables), this method
helps us find least-biased probability distributions subject to the
constraints. These probability distributions are guaranteed to be unique
given our prior knowledge and represent the most we can say about a
system without making more assumptions. For example, if the only thing
we know about a biological community is its average number of
individuals per species, the least-biased inference we could make on its
species abundance distribution is the exponential distribution (Frank
and Smith 2011; Harte and Newman 2014). However, this does not imply
that this distribution will be the best fit to empirical data. The
challenge is to find the right set of constraints that would best
reproduce distributions found in nature.

MaxEnt states that the least-biased probability distribution given the
constraints used is the one with the highest entropy among all
probability distributions that satisfy these constraints. Entropy is a
measure of the average amount of information given by the outcome of a
random variable. Many measures of entropy have been developed in physics
(Beck 2009), but only a fraction of them could be used as an
optimization measure with the principle of maximum entropy. According to
Beck (2009) and Khinchin (2013), a measure of entropy \(H\) should
satisfy four properties in the discrete case: (1) it should be a
function of a probability distribution \(p(n)\) only; (2) it should be
maximized when \(p(n)\) is uniform; (3) it should not be influenced by
outcomes with a null probability; and (4) it should be independent of
the order of information acquisition. The Shannon's entropy (Shannon
1948)

\begin{equation}\protect\hypertarget{eq:shannon}{}{H = -\sum_{n} p(n) \log p(n)}\label{eq:shannon}\end{equation}

satisfies all of these properties. Finding the probability distribution
\(p(n)\) that maximizes \(H\) under a set of \(m\) constraints \(g\) can
be done using the method of Lagrange multipliers. These constraints
could include one or many properties of the probability distribution
(e.g., its mean, variance, and range). However, the normalization
constraint always need to be included in \(g\) in order to make sure
that \(p(n)\) sums to \(1\). The objective is then to find the values of
the Lagrange multipliers \(\lambda_i\) that optimize a function \(F\):

\begin{equation}\protect\hypertarget{eq:F}{}{F = H - \sum_{i=1}^m \lambda_i (g_i-c_i),}\label{eq:F}\end{equation}

where \(g_i\) is the mathematical formulation of the constraint \(i\)
and \(c_i\), its value. Note that \(F\) is just Shannon's entropy to
which we added terms that each sums to zero (\(g_i = c_i\)). \(F\) is
maximized by setting to \(0\) its partial derivative with respect to
\(p(n)\).

The principle of maximum entropy has been used in a wide range of
disciplines, from thermodynamics, chemistry and biology (Martyushev and
Seleznev 2006) to graph and network theory (e.g., Park and Newman 2004;
van der Hoorn, Lippner, and Krioukov 2018). It has also been proven
useful in ecology, e.g.~in species distribution models (Phillips,
Anderson, and Schapire 2006) and macroecological models (Harte et al.
2008; Harte and Newman 2014). In network ecology, MaxEnt has been used
to predict the degree distribution of bipartite networks from the number
of species and the number of interactions (Williams 2011) and to predict
interaction strengths between species pairs using their relative
abundances within an optimal transportation theory regularized with
information entropy (Stock, Poisot, and De Baets 2021). However, to the
best of our knowledge, MaxEnt has never been used to predict food-web
structure directly, even though food webs are among the most documented
and widespread ecological networks (Ings et al. 2009).

Food-web properties that can be derived using MaxEnt are varied and
pertain to different elements of the network (i.e.~at the species
(node), the interaction (edge) or the community (network) levels).
Because MaxEnt is a method of finding least-biased probability
distributions given partial knowledge about a system, these properties
need to be represented probabilistically. For example, at the species
level, MaxEnt can be used to predict the distribution of trophic levels
among species, as well as the distribution of species' vulnerability
(number of predators) and generality (number of prey). By contrast, at
the interaction level, predictions can be made on the distribution of
interaction strengths in weighted food webs. At the community level, it
can generate probability distributions of many measures of their
emerging structure and of networks themselves (i.e.~a probability
distribution that specific network configurations are realized given the
model and constraints). Overall, the potential of this method in the
study of food webs is broad. The applicability and performance of MaxEnt
mostly depend on the ecological information available and on our
capacity to find the right set of state variables that best represent
natural systems and to translate them into appropriate statistical
constraints. Having a validated maximum entropy model for the system at
hand allows us to make least-biased predictions using a minimal amount
of data, as well as identify the most important ecological processes
shaping that system. In other words, MaxEnt can help us better
understand and predict the structure of ecological networks worldwide.

\hypertarget{analytical-and-heuristic-approaches}{%
\subsection{Analytical and heuristic
approaches}\label{analytical-and-heuristic-approaches}}

In this contribution, we used two complementary approaches to predict
the structure of food webs using the principle of maximum entropy. The
first approach consists in deriving constrained probability
distributions of given network properties analytically, whereas the
second approach consists in finding the adjacency matrix of maximum
entropy heuristically, from which network properties can be measured. We
compared our predictions against empirical data and null and neutral
models commonly used in network ecology. We focus on deterministic and
unweighted (Boolean) food webs in both approaches for data availability
reasons. However, our framework can be applied to all types of
ecological networks and a wide variety of measures.

For the first approach (analytic), we focus on species level properties.
Specifically, we derived the joint degree distribution (i.e.~the joint
probability distribution that a species has a given number of prey and
predators in its network) of maximum entropy using only the number of
species \(S\) and the number of interactions \(L\) as state variables.
Then, we predicted the degree distribution of maximum entropy directly
from the joint degree distribution since the first is the sum of the
marginal distributions of the second. Because of the scarcity of
empirical data on the number of interactions in food webs, we present a
method to predict \(L\) from \(S\) (Box 1), thus allowing the prediction
of the joint degree distribution from \(S\) solely.

For the second approach (heuristic), we focus on network level
properties. We used a flexible and heuristic model based on simulated
annealing (an optimization algorithm) to find the network configuration
\emph{close} to maximum entropy and measured its structure. We developed
this heuristic model because the analytical derivation of a maximum
entropy graph model of food webs is difficult, and because this model is
readily applicable to other types of ecological networks and measures.
Indeed, the mathematical representation of food webs (i.e.~directed
simple graphs frequently having self-loops) makes the optimization of
maximum entropy graph models more complicated than with many other types
of (non-ecological) networks. In other words, deriving a probability
distribution on the graphs themselves is difficult when working with
food webs. We built two types of heuristic MaxEnt models depending on
the constraint used. Our type I MaxEnt model uses the connectance of the
network (i.e.~the ratio \(L/S^2\)) as a constraint, whereas our type II
MaxEnt model uses the whole joint degree sequence as a constraint.

\hypertarget{analytical-maximum-entropy-models}{%
\section{Analytical maximum entropy
models}\label{analytical-maximum-entropy-models}}

Our analytical approach is the most common way to use and develop
maximum entropy models. As shown above, starting from a defined set of
constraints, a mathematical expression for the target distribution is
derived using the method of Lagrange multipliers. The derived
distribution of maximum entropy is unique and least biased given the
constraints used. Although we refer to this approach as analytic,
finding the values of the Lagrange multipliers usually requires the use
of numerical methods. Here we use MaxEnt to derive two species level
properties in food webs: the joint degree distribution and the degree
distribution. The degree distribution has driven the attention of
ecologists because of its role in determining the assembly of ecological
networks (Vázquez 2005), shaping their emerging structure (Fortuna et
al. 2010), and understanding interaction mechanisms (Williams 2011). As
noted above, although the degree distribution of maximum entropy has
already been derived in bipartite networks (Williams 2011), we show in
much greater details its mathematical derivation in food webs. But
first, we derive the joint degree distribution, a related property that
holds significantly more ecological information than the degree
distribution.

We tested our analytical MaxEnt model against open food-web data queried
from three different sources and integrated into what we call our
\emph{complete dataset}. These sources include (1) terrestrial and
aquatic food webs sampled globally and archived on the ecological
interactions database Mangal; (2) freshwater stream food webs from New
Zealand; and (3) aquatic food webs from Tuesday Lake (Michigan, United
States). First, all food webs archived on \texttt{mangal.io} (Poisot et
al. 2016; Banville, Vissault, and Poisot 2021) were directly queried
from the database (\(N = 235\)). Most ecological networks archived on
Mangal are multilayer networks, i.e.~networks that describe different
types of interactions. We kept all networks whose interactions were
mainly of predation and herbivory types, and removed the largest network
(\(S = 714\)) for computational efficiency reasons. Then, to this set we
added food webs from two different sources: the New Zealand dataset
(\(N = 21\); J. P. Pomeranz et al. 2018) and the Tuesday Lake dataset
(\(N = 2\); Cohen, Jonsson, and Carpenter 2003). Our complete dataset
thus contained a total of \(257\) food webs. All code and data to
reproduce this article are available at the Open Science Framework
(OSF.IO/KT4GS). Data cleaning, simulations and analyses were conducted
in Julia v1.8.0.

\hypertarget{joint-degree-distribution}{%
\subsection{Joint degree distribution}\label{joint-degree-distribution}}

The joint degree distribution \(p(k_{in},k_{out})\) of a food web with
\(S\) species is a joint discrete probability distribution describing
the probability that a species has \(k_{in}\) predators and \(k_{out}\)
prey, with \(k_{in}\) and \(k_{out}\) \(\epsilon\) \([0, S]\). Basal
species (e.g., plants) have a \(k_{out}\) of \(0\), whereas top
predators have a \(k_{in}\) of \(0\). In contrast, the maximum number of
prey and predators a species can have is set by the number of species in
the food web. Here we show how the joint degree distribution of maximum
entropy can be obtained given knowledge of the number of species \(S\)
and the number of interactions \(L\).

We want to maximize Shannon's entropy

\begin{equation}\protect\hypertarget{eq:entropy_jdd}{}{H = -\sum_{k_{in}=0}^S\sum_{k_{out}=0}^S p(k_{in},k_{out}) \log p(k_{in},k_{out})}\label{eq:entropy_jdd}\end{equation}

subject to the following constraints:

\begin{equation}\protect\hypertarget{eq:g1}{}{g_1 = \sum_{k_{in}=0}^S\sum_{k_{out}=0}^S p(k_{in},k_{out}) = 1;}\label{eq:g1}\end{equation}

\begin{equation}\protect\hypertarget{eq:g2}{}{g_2 = \sum_{k_{in}=0}^S\sum_{k_{out}=0}^S k_{in} p(k_{in},k_{out}) = \langle k_{in} \rangle = \frac{L}{S};}\label{eq:g2}\end{equation}

\begin{equation}\protect\hypertarget{eq:g3}{}{g_3 = \sum_{k_{in}=0}^S\sum_{k_{out}=0}^S k_{out} p(k_{in},k_{out}) = \langle k_{out} \rangle = \frac{L}{S}.}\label{eq:g3}\end{equation}

The first constraint \(g_1\) is our normalizing constraint, whereas the
other two (\(g_2\) and \(g_3\)) fix the average of the marginal
distributions of \(k_{in}\) and \(k_{out}\) to the linkage density
\(L/S\). It is important to notice that
\(\langle k_{in} \rangle = \langle k_{out} \rangle\) because every edge
is associated to a predator and a prey. Therefore, without using any
further constraints, we would expect the joint degree distribution of
maximum entropy to be a symmetric probability distribution with regards
to \(k_{in}\) and \(k_{out}\). However, this does not mean that the
joint degree \emph{sequence} will be symmetric, since the joint degree
sequence is essentially a random realization of its probabilistic
counterpart.

The joint probability distribution of maximum entropy given these
constraints is found using the method of Lagrange multipliers. To do so,
we seek to maximize the following expression:

\begin{equation}\protect\hypertarget{eq:F_jdd}{}{F = H - \lambda_1(g_1-1)-\lambda_2\left( g_2-\frac{L}{S}\right) - \lambda_3 \left( g_3-\frac{L}{S}\right),}\label{eq:F_jdd}\end{equation}

where \(\lambda_1\), \(\lambda_2\), and \(\lambda_3\) are the Lagrange
multipliers. The probability distribution that maximizes entropy is
obtained by finding these values. As pointed out above, \(F\) is just
Shannon's entropy to which we added terms that each sums to zero (our
constraints). \(F\) is maximized by setting to 0 its partial derivative
with respect to \(p(k_{in},k_{out})\). Because the derivative of a
constant is zero, this gives us:

\begin{equation}\protect\hypertarget{eq:lagrange_jdd}{}{\frac{\partial H}{\partial p(k_{in},k_{out})} = \lambda_1 \frac{\partial g_1}{\partial p(k_{in},k_{out})} + \lambda_2 \frac{\partial g_2}{\partial p(k_{in},k_{out})}+ \lambda_3 \frac{\partial g_3}{\partial p(k_{in},k_{out})}.}\label{eq:lagrange_jdd}\end{equation}

Evaluating the partial derivatives with respect to
\(p(k_{in},k_{out})\), we obtain:

\begin{equation}\protect\hypertarget{eq:lagrange2_jdd}{}{-\log p(k_{in},k_{out}) - 1 = \lambda_1 + \lambda_2 k_{in} + \lambda_3 k_{out}.}\label{eq:lagrange2_jdd}\end{equation}

Then, solving eq.~\ref{eq:lagrange2_jdd} for \(p(k_{in},k_{out})\), we
obtain:

\begin{equation}\protect\hypertarget{eq:lagrange3_jdd}{}{p(k_{in},k_{out}) = \frac{e^{-\lambda_2k_{in}-\lambda_3k_{out}}}{Z},}\label{eq:lagrange3_jdd}\end{equation}

where \(Z = e^{1+\lambda_1}\) is called the partition function. The
partition function ensures that probabilities sum to 1 (our
normalization constraint). It can be expressed in terms of \(\lambda_2\)
and \(\lambda_3\) as follows:

\begin{equation}\protect\hypertarget{eq:Z}{}{Z = \sum_{k_{in}=0}^S\sum_{k_{out}=0}^S e^{-\lambda_2k_{in}-\lambda_3k_{out}}.}\label{eq:Z}\end{equation}

After substituting \(p(k_{in},k_{out})\) in eq.~\ref{eq:g2} and
eq.~\ref{eq:g3}, we get a nonlinear system of two equations and two
unknowns:

\begin{equation}\protect\hypertarget{eq:lagrange4_jdd}{}{\frac{1}{Z}\sum_{k_{in}=0}^S\sum_{k_{out}=0}^S k_{in} e^{-\lambda_2k_{in}-\lambda_3k_{out}}  = \frac{L}{S};}\label{eq:lagrange4_jdd}\end{equation}

\begin{equation}\protect\hypertarget{eq:lagrange5_jdd}{}{\frac{1}{Z}\sum_{k_{in}=0}^S\sum_{k_{out}=0}^S k_{out} e^{-\lambda_2k_{in}-\lambda_3k_{out}}  = \frac{L}{S}.}\label{eq:lagrange5_jdd}\end{equation}

We solved eq.~\ref{eq:lagrange4_jdd} and eq.~\ref{eq:lagrange5_jdd}
numerically using the Julia library \texttt{JuMP.jl} v0.21.8 (Dunning,
Huchette, and Lubin 2017). \texttt{JuMP.jl} supports nonlinear
optimization problems by providing exact second derivatives that
increase the accuracy and performance of its solvers. The estimated
values of \(\lambda_2\) and \(\lambda_3\) can be substituted in
eq.~\ref{eq:lagrange3_jdd} to have a more workable expression for the
joint degree distribution.

\begin{figure}
\hypertarget{fig:joint_dd}{%
\centering
\includegraphics{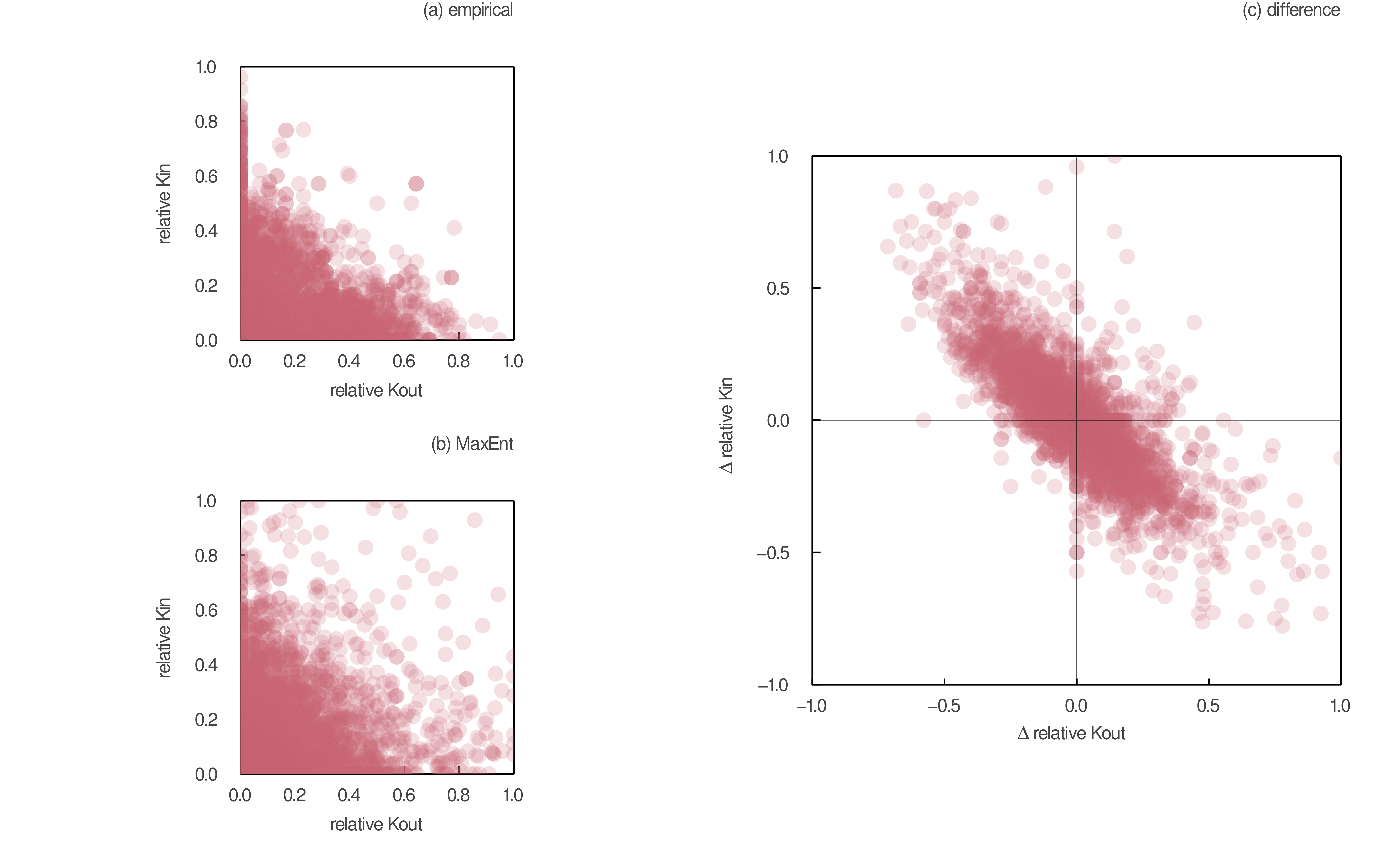}
\caption{Relative number of predators (\(k_{in}\)) as a function of
their relative number of prey (\(k_{out}\)) for each species in (a)
empirical and (b) joint degree sequences obtained from the analytical
MaxEnt model. Empirical networks include most food webs archived on
Mangal, as well as the New Zealand and Tuesday Lake datasets (our
complete dataset). The predicted joint degree sequences were obtained
after sampling one realization of the joint degree distribution of
maximum entropy for each network, while keeping the total number of
interactions constant. (c) Difference between predicted and empirical
values when species are ordered according to their total degrees. In
each panel, each dot corresponds to a single species in one of the
networks.}\label{fig:joint_dd}
}
\end{figure}

We assessed the empirical support of this expression using all food webs
in our complete dataset. First, we predicted the joint degree
distribution of maximum entropy for each of these food webs, i.e.~using
their number of species and number of interactions as state variables.
Then, we sampled one realization of the joint degree sequence for each
network using the probabilities given by the joint degree distribution
of maximum entropy, while fixing the total number of interactions. This
gave us a random realization of the number of prey and predators for
each species in each network. We standardized the predicted \(k_{out}\)
and \(k_{in}\) of each species by the total number of species in their
network to generate relative values, which can be compared across
networks. In fig.~\ref{fig:joint_dd} (left panels), we show the
relationship between these relative \(k_{out}\) and \(k_{in}\) obtained
from the joint degree distributions of maximum entropy (bottom panel)
and this relationship using empirical values (top panel). We observe
that our model predicts higher values of generality and vulnerability
compared to empirical food webs (i.e.~relative values of \(k_{out}\) and
\(k_{in}\) both closer to \(1\)) for many species. In other words, our
model predicts that species that have many predators also have more prey
than what is observed empirically (and conversely). This is not
surprising, given that our model did not include biological factors
preventing generalist predators from having many prey. Nevertheless,
with the exception of these generalist species, MaxEnt adequately
predicts that most species have low generality and vulnerability values.

Examining the difference between predicted and empirical values for each
species gives a slightly different perspective (right panel of
fig.~\ref{fig:joint_dd}). To do so, we must first associate each of our
predictions to a specific species in a network in order to make that
comparison. Indeed, our predicted joint degree sequences have the same
number of species (elements) as their empirical counterparts, but they
are species agnostic. In other words, instead of predicting a value for
each species directly, we predicted the entire joint degree sequence
without taking into account species' identity. The challenge is thus to
adequately associate predictions with empirical data. In
fig.~\ref{fig:joint_dd}, we present these differences when species are
ordered by their total degree in their respective networks (i.e.~by the
sum of their in and out-degrees). This means that the species with the
highest total degree in its network will be associated with the highest
prediction, and so forth. Doing so, we see that species predicted to
have a higher number of predators than what is observed generally have a
lower number of prey than what is observed (and conversely). This is
because the difference in total degree (\(k_{out} + k_{in}\)) between
predictions and empirical data is minimized when species are ranked by
their total degree (i.e.~the average deviation of the sum of relative
\(k_{out}\) and \(k_{in}\) is close to \(0\) across all species). This
result thus shows that the difference between predicted and empirical
total degrees is low for most species when ordered by their total
degrees. There are no apparent biases towards in or out degrees. In
fig.~S1, we show how these differences change when species are instead
ordered by their out-degrees (left panel) and in-degrees (right panel),
respectively.

\begin{figure}
\hypertarget{fig:kl_diverg}{%
\centering
\includegraphics{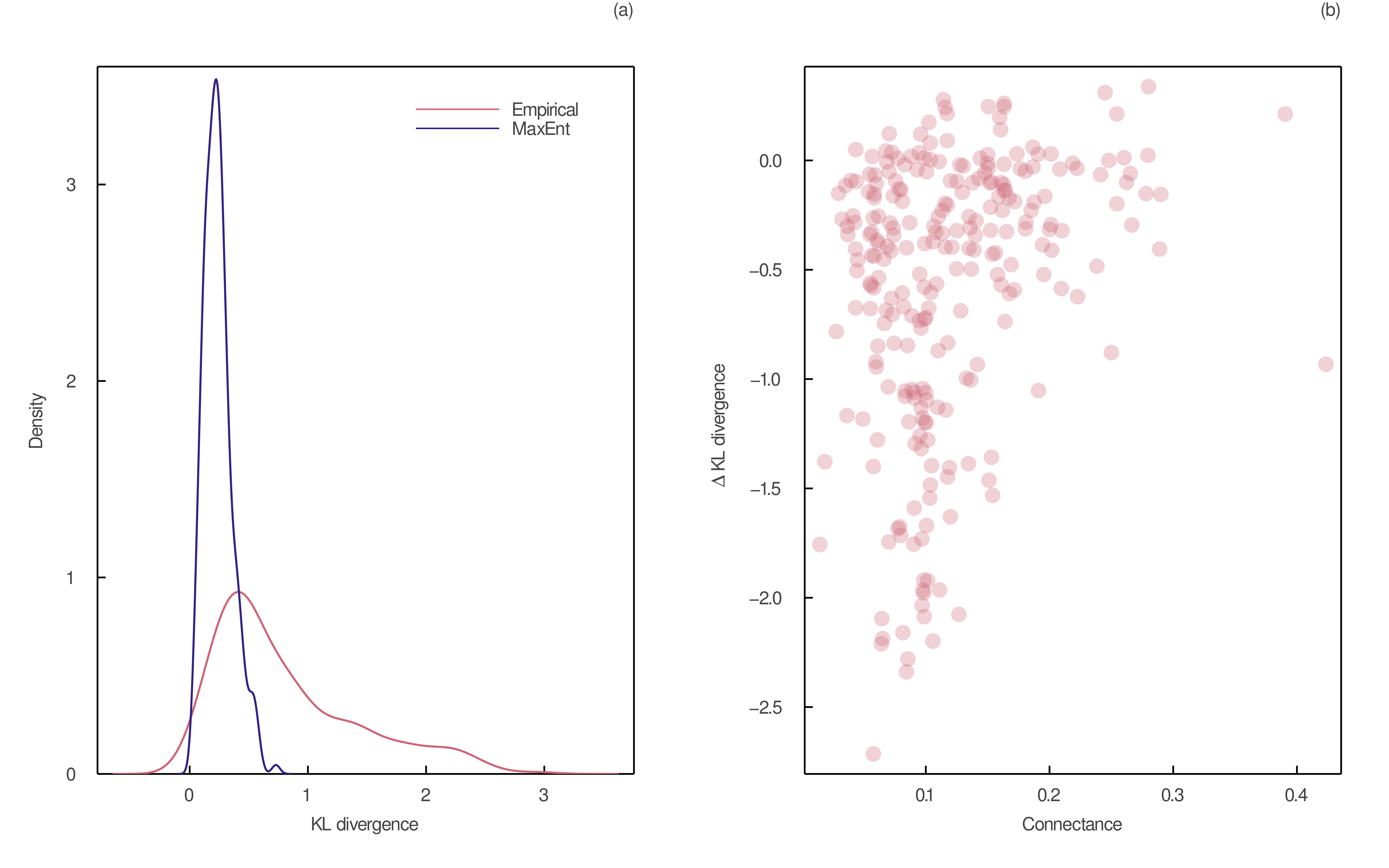}
\caption{(a) Probability density of KL divergence between in and
out-degree sequences of empirical and predicted joint degree sequences.
(b) Difference between the KL divergence of empirical and predicted
joint degree sequences as a function of connectance. In both panels,
empirical networks include most food webs archived on Mangal, as well as
the New Zealand and Tuesday Lake datasets (our complete dataset). The
predicted joint degree sequences were obtained after sampling one
realization of the joint degree distribution of maximum entropy for each
network, while keeping the total number of interactions
constant.}\label{fig:kl_diverg}
}
\end{figure}

Another way to evaluate the empirical support of the predicted joint
degree sequences is to compare their shape with the ones of empirical
food webs. We can describe the shape of a joint degree sequence by
comparing its marginal distributions with one another. To do so, we
calculated the Kullback--Leibler (KL) divergence between the in and
out-degree sequences sampled from the joint degree distribution of
maximum entropy. Similarly, we calculated the divergence between the in
and out-degree sequences obtained empirically. This allows us to compare
the symmetry of empirical and predicted joint degree sequences (left
panel of fig.~\ref{fig:kl_diverg}). As we expected, our model predicts
more similar in-degree and out-degree sequences than empirical data
(shown by lower KL divergence values). However, this difference
decreases with connectance (right panel of fig.~\ref{fig:kl_diverg}).
This might be due to the fact that food webs with a low connectance are
harder to predict than food webs with a high connectance. Indeed, in low
connectance systems, what makes two species interact might be more
important for prediction than in high connectance systems, in which what
prevents species from interacting might be more meaningful. This implies
that more ecological information might be needed in food webs with a low
connectance because more ecological processes determine interactions
compared to non-interactions. Therefore, other ecological constraints
might be needed to account for the asymmetry of the joint degree
distribution, especially for networks with a lower connectance. However,
our MaxEnt model was able to capture quite well the shape of the joint
degree sequence for networks having a high connectance.

\hypertarget{degree-distribution}{%
\subsection{Degree distribution}\label{degree-distribution}}

The joint degree distribution derived using MaxEnt
(eq.~\ref{eq:lagrange3_jdd}) can be used to obtain the degree
distribution of maximum entropy. Indeed, the degree distribution
\(p(k)\) represents the probability that a species has \(k\)
interactions in its food web, with \(k = k_{in} + k_{out}\). It can thus
be obtained from the joint degree distribution as follows:

\[p(k) = \sum_{i=0}^k p(k_{in} = k - i, k_{out} = i).\]

The degree distribution could have also been obtained directly using the
principle of maximum entropy, as discussed in Williams (2011). This
gives the following distribution:

\begin{equation}\protect\hypertarget{eq:lagrange_dd}{}{p(k) = \frac{e^{-\lambda_2k}}{Z},}\label{eq:lagrange_dd}\end{equation}

with \(Z = \sum_{k=0}^S e^{-\lambda_2k}.\)

This can be solved numerically using the constraint of the average
degree \(\langle k \rangle = \frac{2L}{S}\) of a species, yielding an
identical solution to the one obtained using the joint degree
distribution as an intermediate. Note that the mean degree is twice the
value of the linkage density, because every link must be counted twice
when we add in and out-degrees together.

\begin{equation}\protect\hypertarget{eq:lagrange2_dd}{}{\frac{1}{Z}\sum_{k=0}^S k e^{-\lambda_2k} = \frac{2L}{S}}\label{eq:lagrange2_dd}\end{equation}

One aspect of the degree distribution that informs us of its ecological
realism is the number of isolated species it predicts. As MacDonald,
Banville, and Poisot (2020) pointed out, the size of food webs should at
least be of \(S-1\) interactions, since a lower number would yield
isolated species, i.e.~species without any predators or prey. Because
non-basal species must eat to survive, isolated species could indicate
that other species are missing or they could simply be removed from the
food web. In fig.~S2, we show that the degree distribution of maximum
entropy, given \(S\) and \(L\), gives very low probabilities that a
species will be isolated in its food web (i.e.~having \(k = 0\)) above
the \(S-1\) threshold. However, under our purely information-theoretic
model, the probability that a species is isolated is quite high when the
total number of interactions is below \(S-1\). Moreover, the expected
proportion of isolated species rapidly declines by orders of magnitude
with increasing numbers of species and interactions. This supports the
ecological realism of the degree distribution of maximum entropy derived
above. Nevertheless, ecologists wanting to model a system without
allowing isolated species could simply change the lower limit of \(k\)
to \(1\) in eq.~\ref{eq:lagrange2_dd} and solve the resulting equation
numerically.

\hypertarget{box-1---working-with-predicted-numbers-of-interactions}{%
\section{Box 1 - Working with predicted numbers of
interactions}\label{box-1---working-with-predicted-numbers-of-interactions}}

Our analytical MaxEnt models require information on the number of
species and the number of interactions. However, since the later is
rarely measured empirically, ecologists might need to use predictive
models to estimate the total number of interactions in a food web before
using MaxEnt. Here we illustrate how this can be done by combining both
models sequentially.

We used the flexible links model of MacDonald, Banville, and Poisot
(2020) to predict the number of interactions from the number of species.
The flexible links model, in contrast to other predictive models of the
number of interactions, incorporates meaningful ecological constraints
into the prediction of \(L\), namely the minimum \(S-1\) and maximum
\(S^2\) numbers of interactions in food webs. It estimates the
proportion of the \(S^2 - (S - 1)\) \emph{flexible links} that are
realized. More precisely, this model states that the number of
\emph{realized} flexible links (or interactions) \(L_{FL}\) in a food
web represents the number of realized interactions above the minimum
(i.e.~\(L = L_{FL} + S - 1\)) and is obtained from a beta-binomial
distribution with \(S^2 - (S - 1)\) trials and parameters
\(\alpha = \mu e^\phi\) and \(\beta = (1 - \mu) e^\phi\):

\begin{equation}\protect\hypertarget{eq:BB}{}{L_{FL} \sim \mathrm{BB}(S^2 - (S - 1), \mu e^\phi, (1 - \mu) e^\phi),}\label{eq:BB}\end{equation}

where \(\mu\) is the average probability across food webs that a
flexible link is realized, and \(\phi\) is the concentration parameter
around \(\mu\).

We fitted the flexible links model on all food webs in our complete
dataset, and estimated the parameters of eq.~\ref{eq:BB} using a
Hamiltonian Monte Carlo sampler with static trajectory (\(1\) chain and
\(3000\) iterations):

\begin{equation}\protect\hypertarget{eq:BBpost}{}{ [\mu, \phi| \textbf{L}, \textbf{S}] \propto \prod_{i = 1}^{m} \mathrm{BB}(L_i - (S_i - 1) | S_i^2 - (S_i - 1)), \mu e^{\phi}, (1 - \mu) e^\phi) \times \mathrm{B}(\mu| 3 , 7 ) \times \mathcal{N}(\phi | 3, 0.5), }\label{eq:BBpost}\end{equation}

where \(m\) is the number of food webs (\(m = 257\)) and \(\textbf{L}\)
and \(\textbf{S}\) are respectively the vectors of their numbers of
interactions and numbers of species. Our weakly-informative prior
distributions were chosen following MacDonald, Banville, and Poisot
(2020), i.e.~a beta distribution for \(\mu\) and a normal distribution
for \(\phi\). The Monte Carlo sampling of the posterior distribution was
conducted using the Julia library \texttt{Turing} v0.15.12.

\begin{figure}
\hypertarget{fig:degree_dist_fl}{%
\centering
\includegraphics{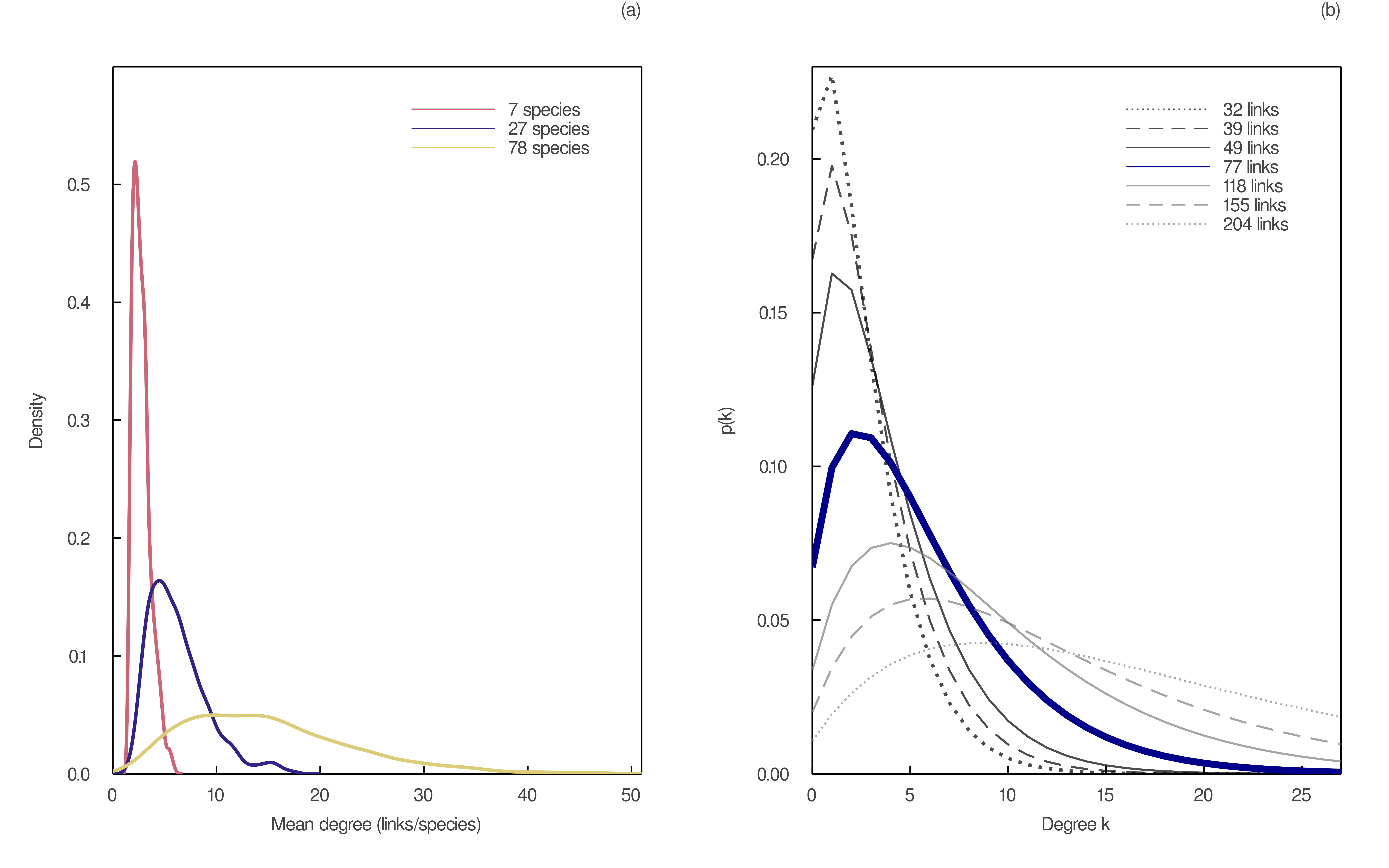}
\caption{(a) Probability density of the mean degree of a food web
obtained using different values of species richness \(S\). The number of
interactions \(L\) was simulated \(1000\) times using the flexible links
model fitted to all empirical networks in our complete dataset
(eq.~\ref{eq:BBpost}). The mean degrees \(2L/S\) were then obtained from
these simulated values. (b) Degree distributions of maximum entropy for
a network of \(S=27\) species and different numbers of interactions. The
numbers of interactions correspond to the lower and upper bounds of the
\(67%
\), \(89%
\), and \(97%
\) percentile intervals (PI), as well as the median (in blue), of the
counterfactuals of the flexible links model. Each degree distribution of
maximum entropy was obtained using eq.~\ref{eq:lagrange_dd} after
solving numerically eq.~\ref{eq:lagrange2_dd} using different values of
the mean degree constraint.}\label{fig:degree_dist_fl}
}
\end{figure}

The flexible links model is a generative model, i.e.~it can generate
plausible values of the predicted variable. We thus simulated \(1000\)
values of \(L\) for different values of \(S\) using the joint posterior
distribution of our model parameters (eq.~\ref{eq:BBpost}), and
calculated the mean degree for each simulated value. The resulting
distributions are shown in the left panel of
fig.~\ref{fig:degree_dist_fl} for three different values of species
richness. In the right panel of fig.~\ref{fig:degree_dist_fl}, we show
how the probability distribution for the mean degree constraints can be
used to generate a distribution of maximum entropy degree distributions,
since each simulated value of mean degree generates a different maximum
entropy degree distribution (eq.~\ref{eq:lagrange_dd} and
eq.~\ref{eq:lagrange2_dd}).

\hypertarget{box-2---corresponding-null-and-neutral-models}{%
\section{Box 2 - Corresponding null and neutral
models}\label{box-2---corresponding-null-and-neutral-models}}

\hypertarget{null-models-types-i-and-ii}{%
\subsection{Null models (types I and
II)}\label{null-models-types-i-and-ii}}

The predictions of our heuristic maximum entropy models were compared
against two topological null models. These null models use the same
ecological information as our heuristic models and thus constitute an
adequate baseline for comparison. The first is the type I null model of
Fortuna and Bascompte (2006), in which the probability that a species
\(i\) predates on another species \(j\) is given by

\begin{equation}\protect\hypertarget{eq:type1null}{}{p(i \rightarrow j) = \frac{L}{S^2}.}\label{eq:type1null}\end{equation}

The second is the type II null model of Bascompte et al. (2003), in
which the probability of interaction is instead given by

\begin{equation}\protect\hypertarget{eq:type2null}{}{p(i \rightarrow j) = \frac{1}{2} \left(\frac{k_{in}(j)}{S} + \frac{k_{out}(i)}{S}\right),}\label{eq:type2null}\end{equation}

where \(k_{in}(j)\) and \(k_{out}(i)\) are the in and out-degrees of
species \(j\) and \(i\), respectively. The type I null model is based on
connectance, whereas the type II null model is based on the joint degree
sequence. Therefore, the type I and II topological null models
correspond to our type I and II heuristic MaxEnt models, respectively,
since they use similar constraints.

We generated probabilistic networks using both types of null models for
all empirical food webs in our complete dataset. Then, we converted
these networks to adjacency matrices of Boolean values by generating
\(100\) random networks for each of these probabilistic webs, and kept
the \(L\) entries that were sampled the most amount of times, with \(L\)
given by the number of interactions in each food web. This ensured that
the resulting null networks had the same number of interactions as their
empirical counterparts. Thus, for each null model, we ended up with one
null adjacency matrix for each empirical network.

\hypertarget{neutral-model}{%
\subsection{Neutral model}\label{neutral-model}}

We also compared our heuristic MaxEnt models with a neutral model of
relative abundances, in which the probabilities of interaction are given
by

\begin{equation}\protect\hypertarget{eq:neutralmodel}{}{p(i \rightarrow j) \propto \frac{n_i}{N} \times \frac{n_j}{N},}\label{eq:neutralmodel}\end{equation}

where \(n_i\) and \(n_j\) are the abundances (or biomass) of both
species, and \(N\) is the total abundance (or biomass) of all species in
the network. We generated neutral abundance matrices for all empirical
food webs in our abundance dataset, and converted these weighted
networks to adjacency matrices of Boolean values using the same method
as the one we used for our null models.

\hypertarget{heuristic-maximum-entropy-models}{%
\section{Heuristic maximum entropy
models}\label{heuristic-maximum-entropy-models}}

With the analytical approach, we showed how important measures of
food-web structure (e.g., the degree distribution and the joint degree
distribution) can be derived with the principle of maximum entropy using
minimal knowledge about a biological community. This type of models,
although useful to make least-biased predictions of many network
properties, can be hard to apply for other measures. Indeed, there are
dozens of measures of network structure (Delmas et al. 2019) and many
are not directly calculated with mathematical equations, but are instead
estimated algorithmically. Moreover, the applicability of this method to
empirical systems is limited by the state variables we can actually
measure and use. Here, we propose a more flexible method to predict many
measures of network structure simultaneously, i.e.~by finding
heuristically the network configuration having maximum entropy given
partial knowledge of its emerging structure.

\hypertarget{from-shannons-to-svd-entropy}{%
\subsection{From Shannon's to SVD
entropy}\label{from-shannons-to-svd-entropy}}

The principle of maximum entropy can be applied on the network itself if
we decompose its adjacency matrix into a non-zero vector of relative
values. This is a necessary step when working with food webs, which are
frequently expressed as a matrix \(A = [a_{ij}]\) of Boolean values
representing the presence (\(a_{ij} = 1\)) or absence (\(a_{ij} = 0\))
of an interaction between two species \(i\) and \(j\). Knowing one or
many properties of a food web of interest (e.g., its number of species
and number of interactions), we can simulate its adjacency matrix
randomly by using these known ecological information to constrain the
space of potential networks. The entropy of this hypothetical matrix can
then be measured after decomposing it into appropriate values.
Simulating a series of networks until we find the one having the highest
entropy allows us to search for the most complex food-web configuration
given the ecological constraints used. This configuration is the least
biased one considering the information available. In other words, the
most we can say about a network's adjacency matrix, without making more
assumptions than the ones given by our incomplete knowledge of its
structure, is the one of maximum entropy. Generating the most complex
network that corresponds to this structure allows us to explore more
easily other properties of food webs under MaxEnt.

Shannon's entropy can only be calculated on conventional probability
distributions such as the joint degree distribution. This is an issue
when working with the adjacency matrix of ecological networks. For this
reason, we need to use another measure of entropy if we want to predict
a network's configuration directly using MaxEnt. We used the SVD entropy
as our measure of entropy, which is an application of Shannon's entropy
to the relative non-zero singular values of a truncated singular value
decomposition (t-SVD; Strydom, Dalla Riva, and Poisot 2021) of a food
web's Boolean adjacency matrix. We measured SVD entropy as follows:

\begin{equation}\protect\hypertarget{eq:svd-entropy}{}{J = -\sum_{i=1}^R s_i \log s_i,}\label{eq:svd-entropy}\end{equation}

where \(s_i\) are the relative singular values of the adjacency matrix
(\(s_i = \sigma_i / \sum_{i = 1}^R \sigma_i\), where \(\sigma_i\) are
the singular values). Note that the distribution of relative singular
values is analogous to a probability distribution, with \(0 < s_i < 1\)
and \(\sum s_i = 1\). This measure also satisfies all four properties of
an appropriate entropy measure above-mentioned, while being a proper
measure of the internal complexity of food webs (Strydom, Dalla Riva,
and Poisot 2021). Following Strydom, Dalla Riva, and Poisot (2021), we
standardized this measure with the rank \(R\) of the matrix
(i.e.~\(J / \ln(R)\)) to account for the difference in dimensions
between networks (\emph{sensu} Pielou's evenness; Pielou 1975).

\hypertarget{types-i-and-ii-heuristic-maxent-models}{%
\subsection{Types I and II heuristic MaxEnt
models}\label{types-i-and-ii-heuristic-maxent-models}}

We used SVD entropy to predict the network configuration of maximum
entropy (i.e.~of maximum complexity) heuristically given different
constraints for all food webs in our complete dataset. We built two
types of heuristic MaxEnt models that differ on the constraint used. The
type I heuristic MaxEnt model is based on connectance, whereas the type
II heuristic MaxEnt model is based on the joint degree sequence. These
models are thus based on the same constraints as the types I (Fortuna
and Bascompte 2006) and II (Bascompte et al. 2003) null models (Box 2)
frequently used to generate random networks topologically. This allows
direct comparison of the performance of null and heuristic MaxEnt models
in reproducing the emerging structure of empirical food webs.

For each network in our complete dataset, we estimated their
configuration of maximum entropy given each of these constraints. For
both types of heuristic MaxEnt models, we used a simulated annealing
algorithm with \(4\) chains, \(2000\) steps and an initial temperature
of \(0.2\). For each food web, we first generated one random Boolean
matrix per chain while fixing the number of species. We also maintained
the total number of interactions (i.e.~the sum of all elements in the
matrix) in the type I MaxEnt model and the joint degree sequence
(i.e.~the rows and columns sums) in the type II MaxEnt model. These were
our initial configurations. Then, we swapped interactions sequentially
while maintaining the original connectance or joint degree sequence.
Configurations with a higher SVD entropy than the previous one in the
chain were always accepted, whereas they were accepted with a
probability conditional to a decreasing temperature and the difference
in SVD entropy when lower. The final configuration with the highest SVD
entropy among the four chains constitute the estimated maximum entropy
configuration of a food web given the constraint used.

\hypertarget{structure-of-maxent-food-webs}{%
\subsection{Structure of MaxEnt food
webs}\label{structure-of-maxent-food-webs}}

We measured various properties of these configurations of maximum
entropy to evaluate how well they reproduce the structure of sampled
food webs. Specifically, we evaluated their nestedness \(\rho\), their
maximum trophic level \(maxtl\), their network diameter \(diam\), their
average maximum similarity between species pairs \(MxSim\) (Williams and
Martinez 2000), their proportion of cannibal species \(Cannib\), their
proportion of omnivorous species \(Omniv\), their SVD entropy, and their
motifs profile. Nestedness indicates how much the diet of specialist
species is a subset of the one of generalists (Delmas et al. 2019) and
was measured using the spectral radius of the adjacency matrix
(Staniczenko, Kopp, and Allesina 2013). In turn, the network diameter
represents the longest of the shortest paths between all species pair
(Albert and Barabasi 2002). Further, cannibal species are species that
can eat individuals of their own species (i.e.~species having self
loops), whereas omnivorous species can prey on different trophic levels
(Williams and Martinez 2000). Finally, a motifs profile represents the
proportion of three-species motifs (subnetworks), which can be
considered as simple building blocks of ecological networks (Milo et al.
2002; Stouffer et al. 2007). All of these properties are relatively easy
to measure and were chosen based on their ecological importance and
prevalent use in network ecology (McCann 2011; Delmas et al. 2019).

We compared the performance of both heuristic MaxEnt models in
predicting these measures to the one of the null and neutral models (Box
2). We conducted these comparisons using two different datasets: (1) our
complete dataset including most food webs archived on Mangal, as well as
all food webs in the New Zealand and Tuesday Lake datasets, and (2) our
\emph{abundance dataset}, a subset of the complete dataset comprising
all food webs having data on their species' relative abundances
(\(N=19\)). Indeed, of the New Zealand and Tuesday Lake datasets, \(19\)
networks had data on species' relative abundances that were used in the
neutral model to better assess the performance of our heuristic models.
We compared our models' predictions using these two datasets separately
to minimize biases and to better represent food webs with abundance data
(tbl.~\ref{tbl:measures_all} and tbl.~\ref{tbl:measures_abund}).

Overall, we found that the models based on the joint degree sequence
(i.e.~the type II null and heuristic MaxEnt models) reproduced the
structure of empirical food webs much better than the ones based on
connectance (i.e.~the type I null and heuristic MaxEnt models). This
suggests that the predictive power of connectance might be more limited
than what was previously suggested (Poisot and Gravel 2014). On the
other hand, the neutral model of relative abundances was surprisingly
good at predicting the maximum trophic level and the network diameter.
However, with the exception of the network diameter, the type II
heuristic MaxEnt model was better at predicting network structure than
the neutral model for most measures considered. This might be because,
although neutral processes are important, they act in concert with niche
processes in determining species interactions Canard et al. (2014). The
joint degree sequence encodes information on both neutral and niche
processes because the number of prey and predators a species has is
determined by its relative abundance and biological traits. These
results thus show that having information on the number of prey and
predators for each species substantially improves the prediction of
food-web structure, both compared to models solely based on connectance
and to the ones solely based on species relative abundances.

Next, the predictions of the type II heuristic MaxEnt model can be
compared to its null model counterpart. On average, the type II
heuristic MaxEnt model was better at predicting nestedness
(\(0.62 \pm 0.08\)) than its corresponding null model
(\(0.73 \pm 0.05\); empirical networks: \(0.63 \pm 0.09\)) for networks
in our complete dataset (tbl.~\ref{tbl:measures_all}). This might in
part be due to the fact that nestedness was calculated using the
spectral radius of the adjacency matrix, which directly leverages
information on the network itself just like the heuristic MaxEnt model.
The proportion of self-loops (cannibal species) was also better
predicted by the type II heuristic MaxEnt model in comparison to the
type II null model. However, the type II null model was better at
predicting network diameter and average maximum similarity between
species pairs, and predictions of the maximum trophic level and the
proportion of omnivorous species were similar between both types of
models. We believe that this is because increasing the complexity of a
food web might increase its average and maximum food-chain lengths. In
comparison, the null model was more stochastic and does not necessarily
produce more complex food webs with longer food-chain lengths.

\hypertarget{tbl:measures_all}{}
\begin{longtable}[]{@{}rrrrrrrr@{}}
\caption{\label{tbl:measures_all}Standardized mean difference between
predicted network measures and empirical data for all food webs in our
complete dataset (\(N = 257\)). Positive (negative) values indicate that
the measure is overestimated (underestimated) on average. Empirical
networks include most food webs archived on Mangal, as well as the New
Zealand and Tuesday Lake food webs. Null 1: Type I null model based on
connectance. MaxEnt 1: Type I heuristic MaxEnt model based on
connectance. Null 2: Type II null model based on the joint degree
sequence. MaxEnt 2: Type II heuristic MaxEnt model based on the joint
degree sequence. \(\rho\): nestedness measured by the spectral radius of
the adjacency matrix. \(maxtl\): maximum trophic level. \(diam\):
network diameter. \(MxSim\): average maximum similarity between species
pairs. \(Cannib\): proportion of cannibal species (self loops).
\(Omniv\): proportion of omnivorous species. \emph{entropy}: SVD
entropy.}\tabularnewline
\toprule
model & rho & maxtl & diam & MxSim & Cannib & Omniv &
entropy\tabularnewline
\midrule
\endfirsthead
\toprule
model & rho & maxtl & diam & MxSim & Cannib & Omniv &
entropy\tabularnewline
\midrule
\endhead
null 1 & -0.167 & 0.980 & 1.428 & -0.502 & 2.007 & 1.493 &
0.056\tabularnewline
MaxEnt 1 & -0.226 & 0.831 & 1.274 & -0.524 & 1.982 & 1.863 &
0.106\tabularnewline
null 2 & 0.160 & -0.125 & 0.016 & 0.007 & 1.078 & 0.559 &
-0.023\tabularnewline
MaxEnt 2 & -0.015 & 0.178 & 0.565 & -0.282 & 0.698 & 0.589 &
0.058\tabularnewline
\bottomrule
\end{longtable}

\hypertarget{tbl:measures_abund}{}
\begin{longtable}[]{@{}rrrrrrrr@{}}
\caption{\label{tbl:measures_abund}Standardized mean difference between
predicted network measures and empirical data for all food webs in our
abundance dataset (\(N = 19\)). Positive (negative) values indicate that
the measure is overestimated (underestimated) on average. Empirical
networks include the New Zealand and Tuesday Lake food webs having
abundance data. Neutral: Neutral model of relative abundances. Null 1:
Type I null model based on connectance. MaxEnt 1: Type I heuristic
MaxEnt model based on connectance. Null 2: Type II null model based on
the joint degree sequence. MaxEnt 2: Type II heuristic MaxEnt model
based on the joint degree sequence. \(\rho\): nestedness measured by the
spectral radius of the adjacency matrix. \(maxtl\): maximum trophic
level. \(diam\): network diameter. \(MxSim\): average maximum similarity
between species pairs. \(Cannib\): proportion of cannibal species (self
loops). \(Omniv\): proportion of omnivorous species. \emph{entropy}: SVD
entropy.}\tabularnewline
\toprule
model & rho & maxtl & diam & MxSim & Cannib & Omniv &
entropy\tabularnewline
\midrule
\endfirsthead
\toprule
model & rho & maxtl & diam & MxSim & Cannib & Omniv &
entropy\tabularnewline
\midrule
\endhead
neutral & 0.367 & -0.090 & 0.027 & 0.266 & 6.870 & 0.576 &
-0.083\tabularnewline
null 1 & -0.134 & 0.950 & 1.919 & -0.369 & 2.077 & 0.614 &
0.068\tabularnewline
MaxEnt 1 & -0.229 & 1.020 & 1.946 & -0.355 & 2.215 & 0.801 &
0.121\tabularnewline
null 2 & 0.128 & -0.115 & -0.135 & 0.157 & 1.444 & 0.029 &
-0.021\tabularnewline
MaxEnt 2 & -0.010 & 0.054 & 0.243 & -0.062 & -0.038 & 0.083 &
0.038\tabularnewline
\bottomrule
\end{longtable}

\begin{figure}
\hypertarget{fig:measures}{%
\centering
\includegraphics{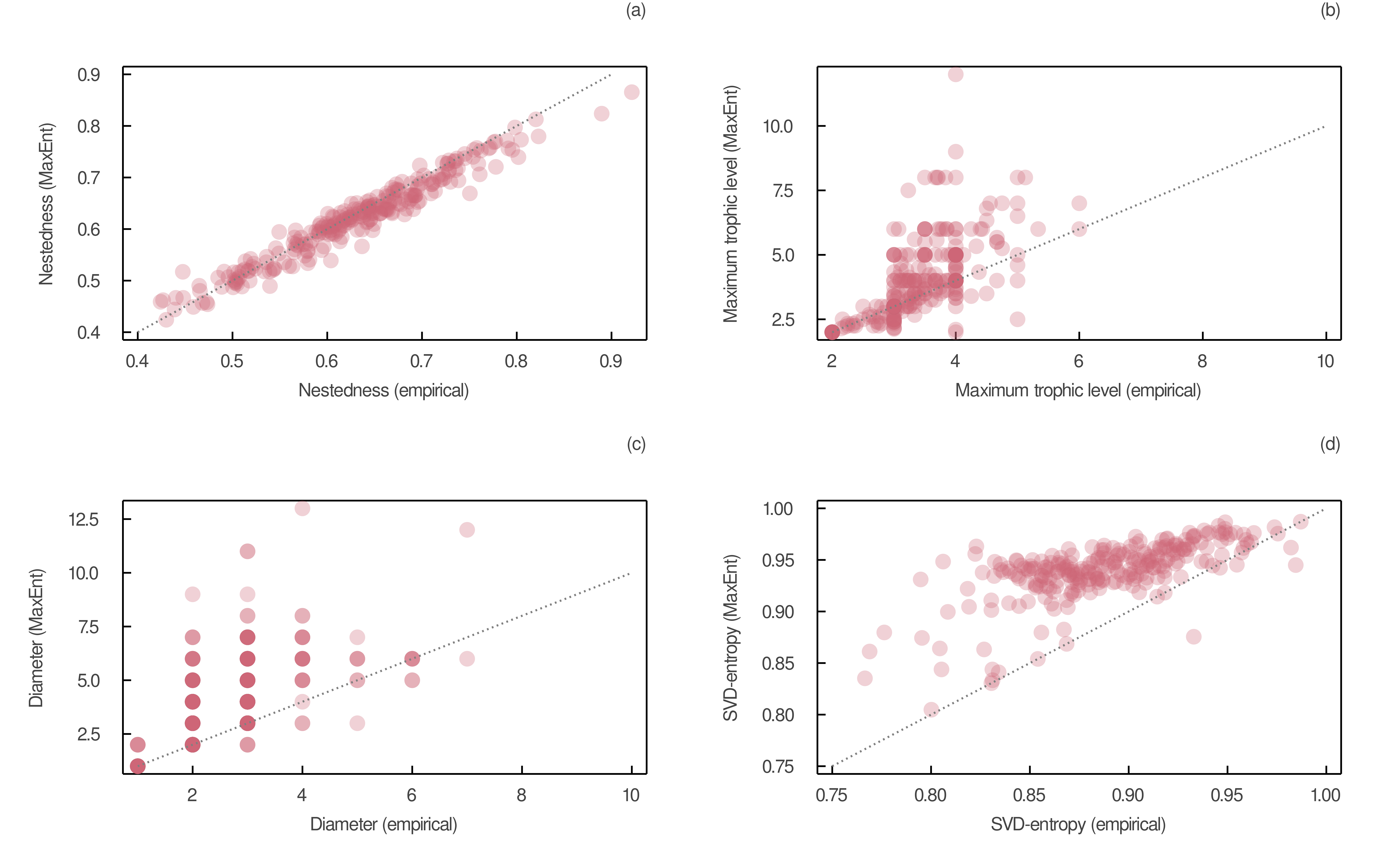}
\caption{Relationship between the structure of empirical and maximum
entropy food webs. Empirical networks include most food webs archived on
Mangal, as well as the New Zealand and Tuesday Lake food webs (our
complete dataset). Maximum entropy networks were obtained using the type
II heuristic MaxEnt model based on the joint degree sequence. (a)
Nestedness (estimated using the spectral radius of the adjacency
matrix), (b) the maximum trophic level, (c) the network diameter, and
(d) the SVD entropy were measured on these empirical and maximum entropy
food webs. The identity line is plotted in each
panel.}\label{fig:measures}
}
\end{figure}

Despite this increase in maximum trophic level and network diameter in
MaxEnt food webs, we found that empirical food webs are close to their
maximum entropy given a fixed joint degree sequence (fig.~S3). Empirical
food webs in the complete dataset had an SVD entropy of
\(0.89 \pm 0.04\), compared to an SVD entropy of \(0.94 \pm 0.03\) for
networks generated using the type II heuristic MaxEnt model. The
relationship between the SVD entropy of empirical food webs and their
maximum entropy is plotted in the last panel of fig.~\ref{fig:measures}.
As expected, the SVD entropy of maximum entropy food webs was higher
than that of empirical food webs for almost all networks, confirming
that our method indeed generated more complex networks. Moreover, we
found no to a weak relationship between the increase in SVD entropy and
the number of species, the number of interactions, and connectance
(fig.~S4). This suggests that the slight increase in entropy between
empirical food webs and their maximum entropy configuration was
sufficient to modify some of their properties, regardless of their
number of species and their number of interactions.

A direct comparison of the structure of maximum entropy food webs,
constrained by the joint degree sequence, with empirical data also
supports the results depicted in tbl.~\ref{tbl:measures_all}. Indeed, in
fig.~\ref{fig:measures} we show how well empirical measures are
predicted by the type II heuristic MaxEnt model. In accordance with our
previous results, we found that nestedness was very well predicted by
our model. However, the model overestimated the maximum trophic level
and network diameter, especially when the sampled food web had
intermediate values of these measures. In fig.~S5, we show that the
pairwise relationships between the four measures in
fig.~\ref{fig:measures} and species richness in empirical food webs are
similar (in magnitude and sign) to the ones found in food webs generated
using the type II heuristic MaxEnt model. This indicates that the number
of species in the network does not seem to impact the ability of the
model to reproduce food-web structure.

\begin{figure}
\hypertarget{fig:motifs}{%
\centering
\includegraphics{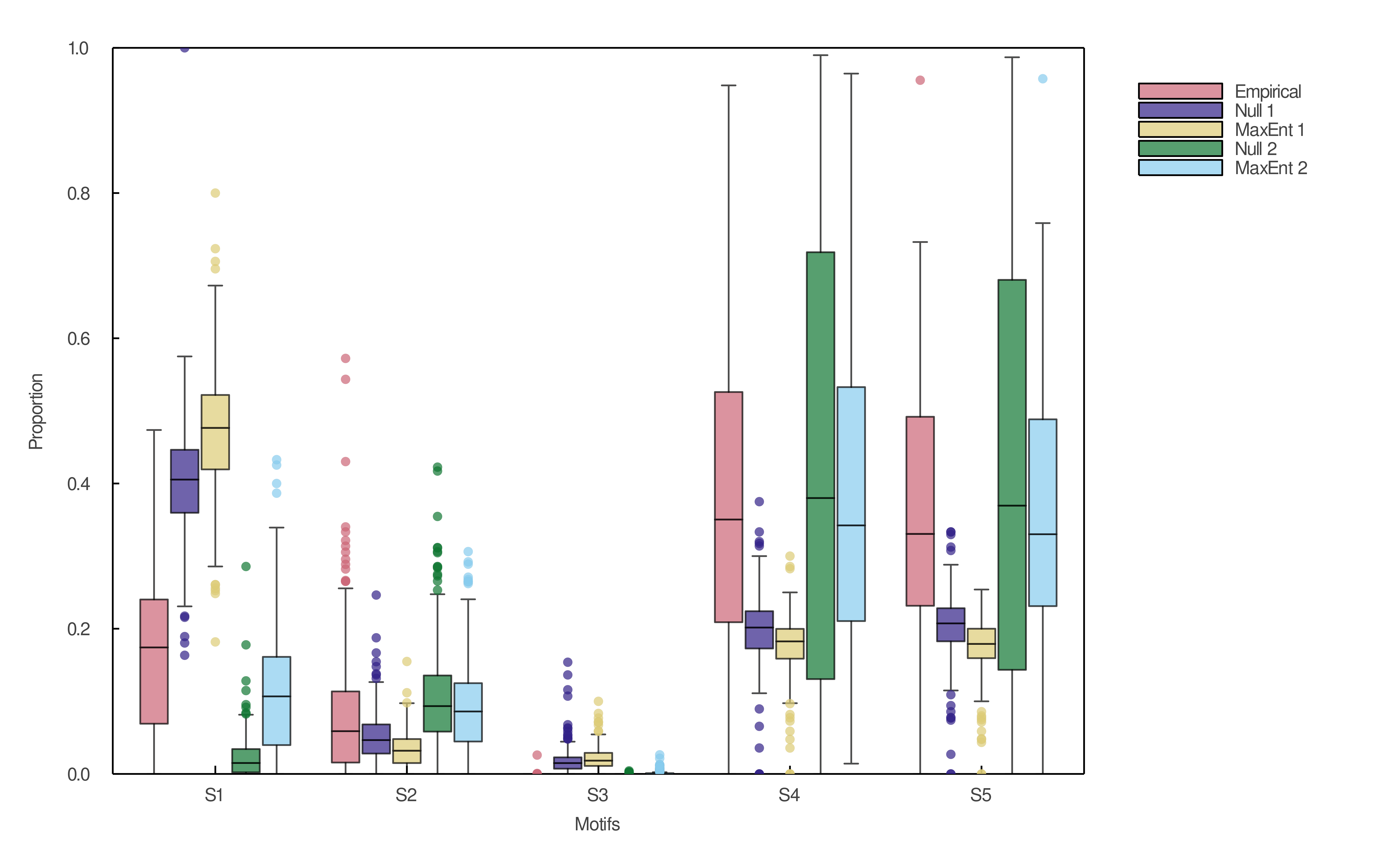}
\caption{Motifs profile of empirical and predicted food webs. Empirical
networks include most food webs archived on Mangal, as well as the New
Zealand and Tuesday Lake datasets (our complete dataset). Null 1: Type I
null model based on connectance. MaxEnt 1: Type I heuristic MaxEnt model
based on connectance. Null 2: Type II null model based on the joint
degree sequence. MaxEnt 2: Type II heuristic MaxEnt model based on the
joint degree sequence. Boxplots display the median proportion of each
motif (middle horizontal lines), as well as the first (bottom horizontal
lines) and third (top horizontal lines) quartiles. Vertical lines
encompass all data points that fall within \(1.5\) times the
interquartile range from both quartiles, and dots are data points that
fall outside this range. Motifs names are from Stouffer et al.
(2007).}\label{fig:motifs}
}
\end{figure}

\begin{figure}
\hypertarget{fig:motifs_rel}{%
\centering
\includegraphics{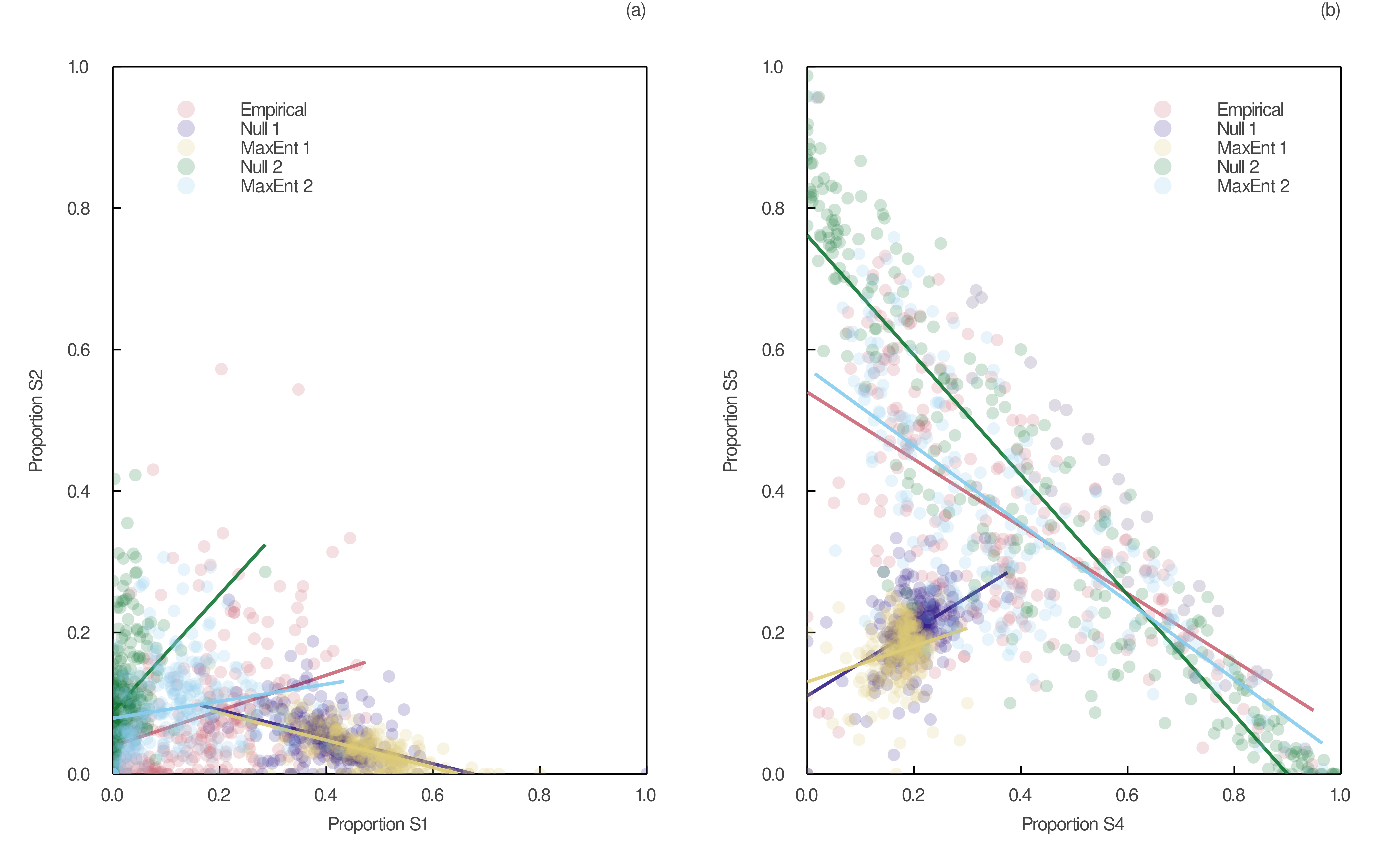}
\caption{Pairwise relationships between motifs proportions of empirical
and predicted food webs. Empirical networks include most food webs
archived on Mangal, as well as the New Zealand and Tuesday Lake datasets
(our complete dataset). Null 1: Type I null model based on connectance.
MaxEnt 1: Type I heuristic MaxEnt model based on connectance. Null 2:
Type II null model based on the joint degree sequence. MaxEnt 2: Type II
heuristic MaxEnt model based on the joint degree sequence. Regression
lines are plotted in each panel. Motifs names are from Stouffer et al.
(2007).}\label{fig:motifs_rel}
}
\end{figure}

Notwithstanding its difficulties in reproducing adequate measures of
food-chain lengths, the type II heuristic MaxEnt model can predict
surprisingly well the motifs profile. Motifs are the backbone of complex
ecological networks from which network structure is built upon and play
a crucial role in community dynamics and assembly (Stouffer and
Bascompte 2011). For this reason, the motifs profile can act as an
effective ecological constraint shaping species interactions networks,
and thus constitute a substantial source of ecological information. In
fig.~\ref{fig:motifs}, we show that the motifs profile of networks
generated using the type II heuristic MaxEnt model accurately reproduced
the one of empirical data. This model made significantly better
predictions than the ones based on connectance and the type II null
model based on the joint degree sequence. This is also shown in
fig.~\ref{fig:motifs_rel}, where we see that the relationships between
motifs proportions in empirical food webs are very similar to the ones
in networks generated using the type II heuristic MaxEnt model. This is
in contrast with the type I null and MaxEnt models based on connectance,
which produced opposite relationships than what was observed
empirically. Our findings thus suggest that increasing food-web
complexity within a maximum entropy framework based on the joint degree
sequence does not alter the proportion of three-species motifs, but
might alter their position relative to one another.

One of the challenges in implementing and validating a maximum entropy
model is to discover where its predictions break down. The results
depicted in tbl.~\ref{tbl:measures_all} and fig.~\ref{fig:measures} show
that our type II heuristic MaxEnt model can capture many high-level
properties of food webs, but does a poor job of capturing others. This
suggests that, although the joint degree sequence is an important driver
of food-web structure, other ecological constraints might be needed to
account for some emerging food-web properties, especially the ones
regarding food-chain lengths. Nevertheless, fig.~\ref{fig:motifs} and
fig.~\ref{fig:motifs_rel} show that the model can reproduce surprisingly
well the motifs profile, one of the most ecologically informative
properties of food webs. This suggests that the emerging structure of
food webs is mainly driven by their joint degree sequence, although
higher-level properties might be needed to ensure that food-chain
lengths fall within realistic values.

\hypertarget{conclusion}{%
\section{Conclusion}\label{conclusion}}

The principle of maximum entropy is a mathematical method of finding
least-biased probability distributions that have some specified
properties given by prior knowledge about a system. We first applied
this conventional MaxEnt approach on food webs to predict species level
properties, namely the joint degree distribution and the degree
distribution of maximum entropy given known numbers of species and
interactions. We found that the joint degree distributions of maximum
entropy had a similar shape to the ones of empirical food webs in
high-connectance systems. However, these MaxEnt distributions were more
symmetric than the ones of empirical food webs when connectance was low,
which suggests that other constraints might be needed to improve these
predictions in low-connectance systems. Then, we used a slightly
different approach that aimed at finding heuristically the network
configuration with the highest SVD entropy, i.e.~whose vector of
relative singular values has maximum entropy. This network of maximum
entropy is the most complex, or random, given the specified structure.
We found that the heuristic maximum entropy model based on connectance
did not predict the structure of sampled food webs very well. However,
the heuristic maximum entropy model based on the entire joint degree
sequence, i.e.~on the number of prey and predators for each species,
gave more convincing results. Indeed, this model reproduced food-web
structure surprisingly well, including the highly informative motifs
profile. Nevertheless, it was not able to predict realistic food-chain
lengths.

Our results bring to the forefront the role of the joint degree
distribution in shaping food-web structure. This echoes the work of
Fortuna et al. (2010), who found that the degree distribution of
ecological networks drive their emerging structure such as their
nestedness and modularity. Network ecologists tend to focus on several
measures of food webs when studying the ecological consequences of their
structure (McCann 2011; Delmas et al. 2019). In fact, following Williams
(2011), we believe there is a lot more ecological information in the
deviation between these properties in empirical systems and in their
maximum entropy configuration given a fixed joint degree sequence.

\hypertarget{alternative-maxent-models}{%
\subsection{Alternative MaxEnt models}\label{alternative-maxent-models}}

In this contribution, we used a method based on simulated annealing to
find the network configuration with the highest SVD entropy while fixing
some aspects of its structure. However, there are different ways to
generate adjacency matrices using MaxEnt. Another technique, also based
on simulated annealing, could begin by generating a food web randomly
with fixed numbers of species and interactions and calculating its joint
degree distribution. Pairs of interactions could then be swapped
sequentially until we minimize the divergence between the calculated
joint degree distribution and the one of maximum entropy obtained
analytically. In that case, this is the entropy of the joint degree
distribution that would be maximized, not the one of the network's
topology. To a certain extent, this method would thus bridge the gap
between the analytical and heuristic approaches presented in this
article. More research is needed to compare the quality of different
methods generating adjacency matrices of food webs using MaxEnt.

Maximum entropy graph models are another type of methods that predict a
distribution of adjacency matrices under soft or hard constraints (e.g.,
Park and Newman 2004; Cimini et al. 2019). Under hard constraints, every
network with a non-zero probability exactly satisfies the constraints on
its structure. This is in contrast with soft constraints, which require
that networks satisfy them on average (i.e.~many networks with a
non-zero probability do not have the exact structure set by the
constraints). Maximum entropy graph models are helpful because they can
provide probability distributions for many network properties by
measuring the structure of all adjacency matrices with a non-zero
probability. However, we consider that our approach based on simulated
annealing is more flexible and more computationally efficient. Indeed,
many measures of food-web structure are hard to translate into
mathematical constraints. Moreover, because food webs are directed
networks that can have self-loops, it makes the mathematical derivation
of maximum entropy graph models difficult. We believe that identifying
heuristically what really constrains the topology of food webs is a
useful first step before attempting to derive the mathematical
formulation of a maximum entropy graph model for food webs.

\hypertarget{applications}{%
\subsection{Applications}\label{applications}}

Our analytical and heuristic models can be applied for different
purposes. First, they could be used to generate first-order
approximations of a network's properties when state variables are known
empirically. For example, knowing the number of species in an ecological
community, we can predict its number of interactions using the flexible
links model and then predict its joint degree distribution with minimal
biases using the principle of maximum entropy. This could prove
particularly useful when predicting network structure at large spatial
scales, subdividing the study area into smaller communities (e.g., grid
cells). Indeed, because species richness and other ecological data are
increasingly abundant (e.g., Dickinson, Zuckerberg, and Bonter 2010),
validated MaxEnt models can be used to respond to a wider range of
macroecological questions regarding food webs.

Second, our analytical model can be used to generate informative priors
in Bayesian analyses of the structure of ecological networks (e.g.,
Cirtwill et al. 2019). Indeed, the probability distribution of maximum
entropy derived using MaxEnt can be used as a prior that can be
constantly updated with novel data. For instance, if we know the number
of species and the number of interactions, we can get the degree
distribution of maximum entropy, as shown in this contribution. The
degree distribution represents the probability that a species can
interact (as a predator or a prey) with a number of other species. Data
on species interactions can be used to update the prior degree
distribution to generate a more accurate posterior distribution, thus
improving our description and understanding of the system.

Third, our analytical and heuristic models can be used to make better
predictions of pairwise species interactions by constraining the space
of feasible networks, as discussed in Strydom et al. (2021). In other
words, we can use the network configuration and/or specific measures of
food-web structure derived using MaxEnt to ensure that our predictions
of interspecific interactions form feasible networks. This means that
the probability that two species interact can be conditional on the
network structure and on the probability of interactions of all other
species pairs. For that purpose, MaxEnt can be used to predict network
structure when other data is lacking.

Finally, our analytical and heuristic models can be used as alternative
null models of ecological networks to better understand and identify the
ecological processes driving food-web structure. Indeed, these
mechanisms can be better described when analysing the deviation of
empirical data from MaxEnt predictions. A strong deviation would
indicate that ecological mechanisms not encoded in the statistical
constraints are at play for the system at hand. If deviations are
systematic, the maximum entropy model might need to be revised to
include appropriate ecological constraints. This revision process helps
us reflect on and identify what really constrains food-web structure.
However, it is important to note here that tangible ecological
mechanisms cannot be directly inferred from statistical distributions
(Warren II, Costa, and Bradford 2022). Instead, by identifying the
constraints of a system and by analysing empirical deviations from
maximum entropy predictions, MaxEnt can only help us redirect research
efforts towards understanding the biological mechanisms behind these
constraints.

The principle of maximum entropy can thus be applied for both the
prediction and understanding of natural systems. Therefore, the model
interpretation depends on how we use it. It can be used as a baseline
distribution to identify the ecological constraints organizing natural
systems. It can also be used as predictions of ecological systems. This
distinction between understanding and predicting is essential when using
and interpreting MaxEnt models.

\hypertarget{final-remarks}{%
\subsection{Final remarks}\label{final-remarks}}

One of the biggest challenges in using the principle of maximum entropy
is to identify the set of state variables that best reproduce empirical
data. We found that the number of species and the number of interactions
are important state variables for the prediction of the joint degree
distribution. Similarly, we found that the numbers of prey and predators
for each species in a food web are important state variables for the
prediction of the network configuration. However, our predictions
overestimated the symmetry of the joint degree distribution for our
analytical model and the maximum trophic level and network diameter for
our heuristic model. We should thus continue to play the ecological
detective to find these other topological constraints that would improve
the predictions of MaxEnt models and help us understand better what
really drives food-web structure.

\hypertarget{acknowledgments}{%
\section{Acknowledgments}\label{acknowledgments}}

We acknowledge that this study was conducted on land within the
traditional unceded territory of the Saint Lawrence Iroquoian,
Anishinabewaki, Mohawk, Huron-Wendat, and Omàmiwininiwak nations. This
work was supported by the Institute for Data Valorisation (IVADO) and
the Natural Sciences and Engineering Research Council of Canada (NSERC)
Collaborative Research and Training Experience (CREATE) program, through
the Computational Biodiversity Science and Services (BIOS\(^2\))
program.

\hypertarget{references}{%
\section*{References}\label{references}}
\addcontentsline{toc}{section}{References}

\hypertarget{refs}{}
\begin{CSLReferences}{1}{0}
\leavevmode\hypertarget{ref-Albert2002StaMec}{}%
Albert, R., and A. L. Barabasi. 2002. {``Statistical Mechanics of
Complex Networks.''} \emph{Reviews of Modern Physics} 74 (1): 47--97.
\url{https://doi.org/10.1103/RevModPhys.74.47}.

\leavevmode\hypertarget{ref-Banville2021ManJla}{}%
Banville, Francis, Steve Vissault, and Timothée Poisot. 2021.
{``Mangal.jl and EcologicalNetworks.jl: Two Complementary Packages for
Analyzing Ecological Networks in Julia.''} \emph{Journal of Open Source
Software} 6 (61): 2721. \url{https://doi.org/10.21105/joss.02721}.

\leavevmode\hypertarget{ref-Bartomeus2016ComFra}{}%
Bartomeus, Ignasi, Dominique Gravel, Jason M. Tylianakis, Marcelo A.
Aizen, Ian A. Dickie, and Maud Bernard-Verdier. 2016. {``A Common
Framework for Identifying Linkage Rules Across Different Types of
Interactions.''} \emph{Functional Ecology} 30 (12): 1894--1903.
\url{https://doi.org/10.1111/1365-2435.12666}.

\leavevmode\hypertarget{ref-Bascompte2003NesAssa}{}%
Bascompte, J., P. Jordano, C. J. Melian, and J. M. Olesen. 2003. {``The
Nested Assembly of Plant-Animal Mutualistic Networks.''}
\emph{Proceedings of the National Academy of Sciences of the United
States of America} 100 (16): 9383--87.
\url{https://doi.org/10.1073/pnas.1633576100}.

\leavevmode\hypertarget{ref-Beck2009GenInf}{}%
Beck, Christian. 2009. {``Generalised Information and Entropy Measures
in Physics.''} \emph{Contemporary Physics} 50 (4): 495--510.
\url{https://doi.org/10.1080/00107510902823517}.

\leavevmode\hypertarget{ref-Canard2014EmpEva}{}%
Canard, E. F., N. Mouquet, D. Mouillot, M. Stanko, D. Miklisova, and D.
Gravel. 2014. {``Empirical Evaluation of Neutral Interactions in
Host-Parasite Networks.''} \emph{The American Naturalist} 183 (4):
468--79. \url{https://doi.org/10.1086/675363}.

\leavevmode\hypertarget{ref-Cimini2019StaPhy}{}%
Cimini, Giulio, Tiziano Squartini, Fabio Saracco, Diego Garlaschelli,
Andrea Gabrielli, and Guido Caldarelli. 2019. {``The Statistical Physics
of Real-World Networks.''} \emph{Nature Reviews Physics} 1 (1): 58--71.
\url{https://doi.org/10.1038/s42254-018-0002-6}.

\leavevmode\hypertarget{ref-Cirtwill2019QuaFrac}{}%
Cirtwill, Alyssa R., Anna Eklof, Tomas Roslin, Kate Wootton, and
Dominique Gravel. 2019. {``A Quantitative Framework for Investigating
the Reliability of Empirical Network Construction.''} \emph{Methods in
Ecology and Evolution} 10 (6): 902--11.
\url{https://doi.org/10.1111/2041-210X.13180}.

\leavevmode\hypertarget{ref-Cohen2003EcoComa}{}%
Cohen, Joel E., Tomas Jonsson, and Stephen R. Carpenter. 2003.
{``Ecological Community Description Using the Food Web, Species
Abundance, and Body Size.''} \emph{Proceedings of the National Academy
of Sciences} 100 (4): 1781--86.
\url{https://doi.org/10.1073/pnas.232715699}.

\leavevmode\hypertarget{ref-Delmas2019AnaEco}{}%
Delmas, Eva, Mathilde Besson, Marie Hélène Brice, Laura A. Burkle,
Giulio V. Dalla Riva, Marie Josée Fortin, Dominique Gravel, et al. 2019.
{``Analysing Ecological Networks of Species Interactions.''}
\emph{Biological Reviews}. \url{https://doi.org/10.1111/brv.12433}.

\leavevmode\hypertarget{ref-Dickinson2010CitSci}{}%
Dickinson, Janis L., Benjamin Zuckerberg, and David N. Bonter. 2010.
{``Citizen Science as an Ecological Research Tool: Challenges and
Benefits.''} In \emph{Annual Review of Ecology, Evolution, and
Systematics, Vol 41}, edited by D. J. Futuyma, H. B. Shafer, and D.
Simberloff, 41:149--72. Palo Alto: Annual Reviews.
\url{https://doi.org/10.1146/annurev-ecolsys-102209-144636}.

\leavevmode\hypertarget{ref-Dunne2002NetStrb}{}%
Dunne, Jennifer A., Richard J. Williams, and Neo D. Martinez. 2002.
{``Network Structure and Biodiversity Loss in Food Webs: Robustness
Increases with Connectance.''} \emph{Ecology Letters} 5 (4): 558--67.
\url{https://doi.org/10.1046/j.1461-0248.2002.00354.x}.

\leavevmode\hypertarget{ref-Dunning2017JumMod}{}%
Dunning, Iain, Joey Huchette, and Miles Lubin. 2017. {``JuMP: A Modeling
Language for Mathematical Optimization.''} \emph{SIAM Review} 59 (2):
295--320. \url{https://doi.org/10.1137/15M1020575}.

\leavevmode\hypertarget{ref-Fortuna2006HabLos}{}%
Fortuna, M. A., and J. Bascompte. 2006. {``Habitat Loss and the
Structure of Plant-Animal Mutualistic Networks.''} \emph{Ecology
Letters} 9 (3): 278--83.
\url{https://doi.org/10.1111/j.1461-0248.2005.00868.x}.

\leavevmode\hypertarget{ref-Fortuna2010NesMod}{}%
Fortuna, M. A., Daniel B. Stouffer, Jens M. Olesen, Pedro Jordano, David
Mouillot, Boris R. Krasnov, Robert Poulin, and Jordi Bascompte. 2010.
{``Nestedness Versus Modularity in Ecological Networks: Two Sides of the
Same Coin?''} \emph{Journal of Animal Ecology} 79 (4): 811--17.
\url{https://doi.org/10.1111/j.1365-2656.2010.01688.x}.

\leavevmode\hypertarget{ref-Frank2011SimDera}{}%
Frank, S. A., and E. Smith. 2011. {``A Simple Derivation and
Classification of Common Probability Distributions Based on Information
Symmetry and Measurement Scale.''} \emph{Journal of Evolutionary
Biology} 24 (3): 469--84.
\url{https://doi.org/10.1111/j.1420-9101.2010.02204.x}.

\leavevmode\hypertarget{ref-Gomez2011FunConb}{}%
Gómez, José M., Francisco Perfectti, and Pedro Jordano. 2011. {``The
Functional Consequences of Mutualistic Network Architecture.''}
\emph{PLOS ONE} 6 (1): e16143.
\url{https://doi.org/10.1371/journal.pone.0016143}.

\leavevmode\hypertarget{ref-Harremoes2001MaxEnt}{}%
Harremoës, Peter, and Flemming Topsøe. 2001. {``Maximum Entropy
Fundamentals.''} \emph{Entropy} 3 (3): 191--226.
\url{https://doi.org/10.3390/e3030191}.

\leavevmode\hypertarget{ref-Harte2014MaxInf}{}%
Harte, J., and Erica A. Newman. 2014. {``Maximum Information Entropy: A
Foundation for Ecological Theory.''} \emph{Trends in Ecology \&
Evolution} 29 (7): 384--89.
\url{https://doi.org/10.1016/j.tree.2014.04.009}.

\leavevmode\hypertarget{ref-Harte2008MaxEnt}{}%
Harte, J., T. Zillio, E. Conlisk, and A. B. Smith. 2008. {``Maximum
Entropy and the State-Variable Approach to Macroecology.''}
\emph{Ecology} 89 (10): 2700--2711.
\url{https://doi.org/10.1890/07-1369.1}.

\leavevmode\hypertarget{ref-vanderHoorn2018SpaMaxa}{}%
Hoorn, Pim van der, Gabor Lippner, and Dmitri Krioukov. 2018. {``Sparse
Maximum-Entropy Random Graphs with a Given Power-Law Degree
Distribution.''} \emph{Journal of Statistical Physics} 173 (3): 806--44.
\url{https://doi.org/10.1007/s10955-017-1887-7}.

\leavevmode\hypertarget{ref-Hortal2015SevSho}{}%
Hortal, Joaquín, Francesco de Bello, José Alexandre F. Diniz-Filho,
Thomas M. Lewinsohn, Jorge M. Lobo, and Richard J. Ladle. 2015. {``Seven
Shortfalls That Beset Large-Scale Knowledge of Biodiversity.''}
\emph{Annual Review of Ecology, Evolution, and Systematics} 46 (1):
523--49. \url{https://doi.org/10.1146/annurev-ecolsys-112414-054400}.

\leavevmode\hypertarget{ref-Ings2009RevEco}{}%
Ings, Thomas C., José M. Montoya, Jordi Bascompte, Nico Blüthgen, Lee
Brown, Carsten F. Dormann, François Edwards, et al. 2009. {``Review:
Ecological Networks Beyond Food Webs.''} \emph{Journal of Animal
Ecology} 78 (1): 253--69.
\url{https://doi.org/10.1111/j.1365-2656.2008.01460.x}.

\leavevmode\hypertarget{ref-Jaynes1957InfThe}{}%
Jaynes, E. T. 1957a. {``Information Theory and Statistical Mechanics.''}
\emph{Physical Review} 106 (4): 620--30.
\url{https://doi.org/10.1103/PhysRev.106.620}.

\leavevmode\hypertarget{ref-Jaynes1957InfThea}{}%
---------. 1957b. {``Information Theory and Statistical Mechanics.
II.''} \emph{Physical Review} 108 (2): 171--90.
\url{https://doi.org/10.1103/PhysRev.108.171}.

\leavevmode\hypertarget{ref-Jordano2016SamNeta}{}%
Jordano, Pedro. 2016. {``Sampling Networks of Ecological
Interactions.''} \emph{Functional Ecology} 30 (12): 1883--93.
\url{https://doi.org/10.1111/1365-2435.12763}.

\leavevmode\hypertarget{ref-Khinchin2013MatFou}{}%
Khinchin, A. Ya. 2013. \emph{Mathematical Foundations of Information
Theory}. Courier Corporation.

\leavevmode\hypertarget{ref-MacDonald2020RevLin}{}%
MacDonald, Arthur Andrew Meahan, Francis Banville, and Timothée Poisot.
2020. {``Revisiting the Links-Species Scaling Relationship in Food
Webs.''} \emph{Patterns} 0 (0).
\url{https://doi.org/10.1016/j.patter.2020.100079}.

\leavevmode\hypertarget{ref-Martyushev2006MaxEnt}{}%
Martyushev, L. M., and V. D. Seleznev. 2006. {``Maximum Entropy
Production Principle in Physics, Chemistry and Biology.''} \emph{Physics
Reports-Review Section of Physics Letters} 426 (1): 1--45.
\url{https://doi.org/10.1016/j.physrep.2005.12.001}.

\leavevmode\hypertarget{ref-McCann2011FooWeb}{}%
McCann, Kevin S. 2011. \emph{Food Webs (MPB-50)}. \emph{Food Webs
(MPB-50)}. Princeton University Press.
\url{https://doi.org/10.1515/9781400840687}.

\leavevmode\hypertarget{ref-Milo2002NetMot}{}%
Milo, R., S. Shen-Orr, S. Itzkovitz, N. Kashtan, D. Chklovskii, and U.
Alon. 2002. {``Network Motifs: Simple Building Blocks of Complex
Networks.''} \emph{Science} 298 (5594): 824--27.
\url{https://doi.org/10.1126/science.298.5594.824}.

\leavevmode\hypertarget{ref-Park2004StaMeca}{}%
Park, Juyong, and M. E. J. Newman. 2004. {``Statistical Mechanics of
Networks.''} \emph{Physical Review E} 70 (6): 066117.
\url{https://doi.org/10.1103/PhysRevE.70.066117}.

\leavevmode\hypertarget{ref-Pascual2006EcoNeta}{}%
Pascual, Department of Ecology and Evolutionary Biology Mercedes, and
Visiting Professor Jennifer A. Dunne. 2006. \emph{Ecological Networks:
Linking Structure to Dynamics in Food Webs}. Oxford University Press,
USA.

\leavevmode\hypertarget{ref-Phillips2006MaxEnta}{}%
Phillips, Steven J., Robert P. Anderson, and Robert E. Schapire. 2006.
{``Maximum Entropy Modeling of Species Geographic Distributions.''}
\emph{Ecological Modelling} 190 (3): 231--59.
\url{https://doi.org/10.1016/j.ecolmodel.2005.03.026}.

\leavevmode\hypertarget{ref-Pielou1975EcoDiv}{}%
Pielou, Evelyn C. 1975. {``Ecological Diversity.''} In. 574.524018 P5.

\leavevmode\hypertarget{ref-Poisot2016ManMakb}{}%
Poisot, Timothée, Benjamin Baiser, Jennifer A. Dunne, Sonia Kefi,
Francois Massol, Nicolas Mouquet, Tamara N. Romanuk, Daniel B. Stouffer,
Spencer A. Wood, and Dominique Gravel. 2016. {``Mangal - Making
Ecological Network Analysis Simple.''} \emph{Ecography} 39 (4): 384--90.
\url{https://doi.org/10.1111/ecog.00976}.

\leavevmode\hypertarget{ref-Poisot2021GloKno}{}%
Poisot, Timothée, Gabriel Bergeron, Kevin Cazelles, Tad Dallas,
Dominique Gravel, Andrew MacDonald, Benjamin Mercier, Clément Violet,
and Steve Vissault. 2021. {``Global Knowledge Gaps in Species
Interaction Networks Data.''} \emph{Journal of Biogeography} 48 (7):
1552--63. \url{https://doi.org/10.1111/jbi.14127}.

\leavevmode\hypertarget{ref-Poisot2014WheEcoa}{}%
Poisot, Timothée, and Dominique Gravel. 2014. {``When Is an Ecological
Network Complex? Connectance Drives Degree Distribution and Emerging
Network Properties.''} \emph{PeerJ} 2: e251.
\url{https://doi.org/10.7717/peerj.251}.

\leavevmode\hypertarget{ref-Poisot2015SpeWhya}{}%
Poisot, Timothée, Daniel B. Stouffer, and Dominique Gravel. 2015.
{``Beyond Species: Why Ecological Interaction Networks Vary Through
Space and Time.''} \emph{Oikos} 124 (3): 243--51.
\url{https://doi.org/10.1111/oik.01719}.

\leavevmode\hypertarget{ref-Pomeranz2019InfPrea}{}%
Pomeranz, Justin P. F., Ross M. Thompson, Timothée Poisot, and Jon S.
Harding. 2019. {``Inferring Predatorprey Interactions in Food Webs.''}
\emph{Methods in Ecology and Evolution} 10 (3): 356--67.
\url{https://doi.org/10.1111/2041-210X.13125}.

\leavevmode\hypertarget{ref-Pomeranz2018DatInf}{}%
Pomeranz, Justin Page, Ross M. Thompson, Timothée Poisot, Jon S.
Harding, and Justin P. F. Pomeranz. 2018. {``Data from: Inferring
Predator-Prey Interactions in Food Webs.''} Dryad.
\url{https://doi.org/10.5061/DRYAD.K59M37F}.

\leavevmode\hypertarget{ref-Shannon1948MatThe}{}%
Shannon, C. E. 1948. {``A Mathematical Theory of Communication.''}
\emph{The Bell System Technical Journal} 27 (3): 379--423.
\url{https://doi.org/10.1002/j.1538-7305.1948.tb01338.x}.

\leavevmode\hypertarget{ref-Staniczenko2013GhoNes}{}%
Staniczenko, Phillip P. A., Jason C. Kopp, and Stefano Allesina. 2013.
{``The Ghost of Nestedness in Ecological Networks.''} \emph{Nature
Communications} 4 (1): 1391. \url{https://doi.org/10.1038/ncomms2422}.

\leavevmode\hypertarget{ref-Stock2021OptTra}{}%
Stock, Michiel, Timothée Poisot, and Bernard De Baets. 2021. {``Optimal
Transportation Theory for Species Interaction Networks.''} \emph{Ecology
and Evolution} 11 (9): 3841--55.
\url{https://doi.org/10.1002/ece3.7254}.

\leavevmode\hypertarget{ref-Stouffer2011ComInc}{}%
Stouffer, Daniel B., and Jordi Bascompte. 2011. {``Compartmentalization
Increases Food-Web Persistence.''} \emph{Proceedings of the National
Academy of Sciences} 108 (9): 3648--52.
\url{https://doi.org/10.1073/pnas.1014353108}.

\leavevmode\hypertarget{ref-Stouffer2007EviExi}{}%
Stouffer, Daniel B., Juan Camacho, Wenxin Jiang, and Luis A. Nunes
Amaral. 2007. {``Evidence for the Existence of a Robust Pattern of Prey
Selection in Food Webs.''} \emph{Proceedings of the Royal Society
B-Biological Sciences} 274 (1621): 1931--40.
\url{https://doi.org/10.1098/rspb.2007.0571}.

\leavevmode\hypertarget{ref-Strydom2021RoaPre}{}%
Strydom, Tanya, Michael D. Catchen, Francis Banville, Dominique Caron,
Gabriel Dansereau, Philippe Desjardins-Proulx, Norma R. Forero-Muñoz, et
al. 2021. {``A Roadmap Towards Predicting Species Interaction Networks
(across Space and Time).''} \emph{Philosophical Transactions of the
Royal Society B: Biological Sciences} 376 (1837): 20210063.
\url{https://doi.org/10.1098/rstb.2021.0063}.

\leavevmode\hypertarget{ref-Strydom2021SvdEnt}{}%
Strydom, Tanya, Giulio V. Dalla Riva, and Timothee Poisot. 2021. {``SVD
Entropy Reveals the High Complexity of Ecological Networks.''}
\emph{Frontiers in Ecology and Evolution} 9: 623141.
\url{https://doi.org/10.3389/fevo.2021.623141}.

\leavevmode\hypertarget{ref-Vazquez2005DegDisa}{}%
Vázquez, Diego P. 2005. {``Degree Distribution in Plantanimal
Mutualistic Networks: Forbidden Links or Random Interactions?''}
\emph{Oikos} 108 (2): 421--26.
\url{https://doi.org/10.1111/j.0030-1299.2005.13619.x}.

\leavevmode\hypertarget{ref-WarrenII2022SeeSha}{}%
Warren II, Robert J., James T. Costa, and Mark A. Bradford. 2022.
{``Seeing Shapes in Clouds: The Fallacy of Deriving Ecological
Hypotheses from Statistical Distributions.''} \emph{Oikos} 2022 (11):
e09315. \url{https://doi.org/10.1111/oik.09315}.

\leavevmode\hypertarget{ref-Williams2011BioMet}{}%
Williams, Richard J. 2011. {``Biology, Methodology or Chance? The Degree
Distributions of Bipartite Ecological Networks.''} \emph{PLOS ONE} 6
(3): e17645. \url{https://doi.org/10.1371/journal.pone.0017645}.

\leavevmode\hypertarget{ref-Williams2000SimRul}{}%
Williams, Richard J., and Neo D. Martinez. 2000. {``Simple Rules Yield
Complex Food Webs.''} \emph{Nature} 404 (6774): 180--83.
\url{https://doi.org/10.1038/35004572}.

\end{CSLReferences}

\end{document}


\hypertarget{supplementary-materials}{%
\section{Supplementary materials}\label{supplementary-materials}}

\begin{figure}
\hypertarget{fig:kin_kout_diff}{%
\centering
\includegraphics{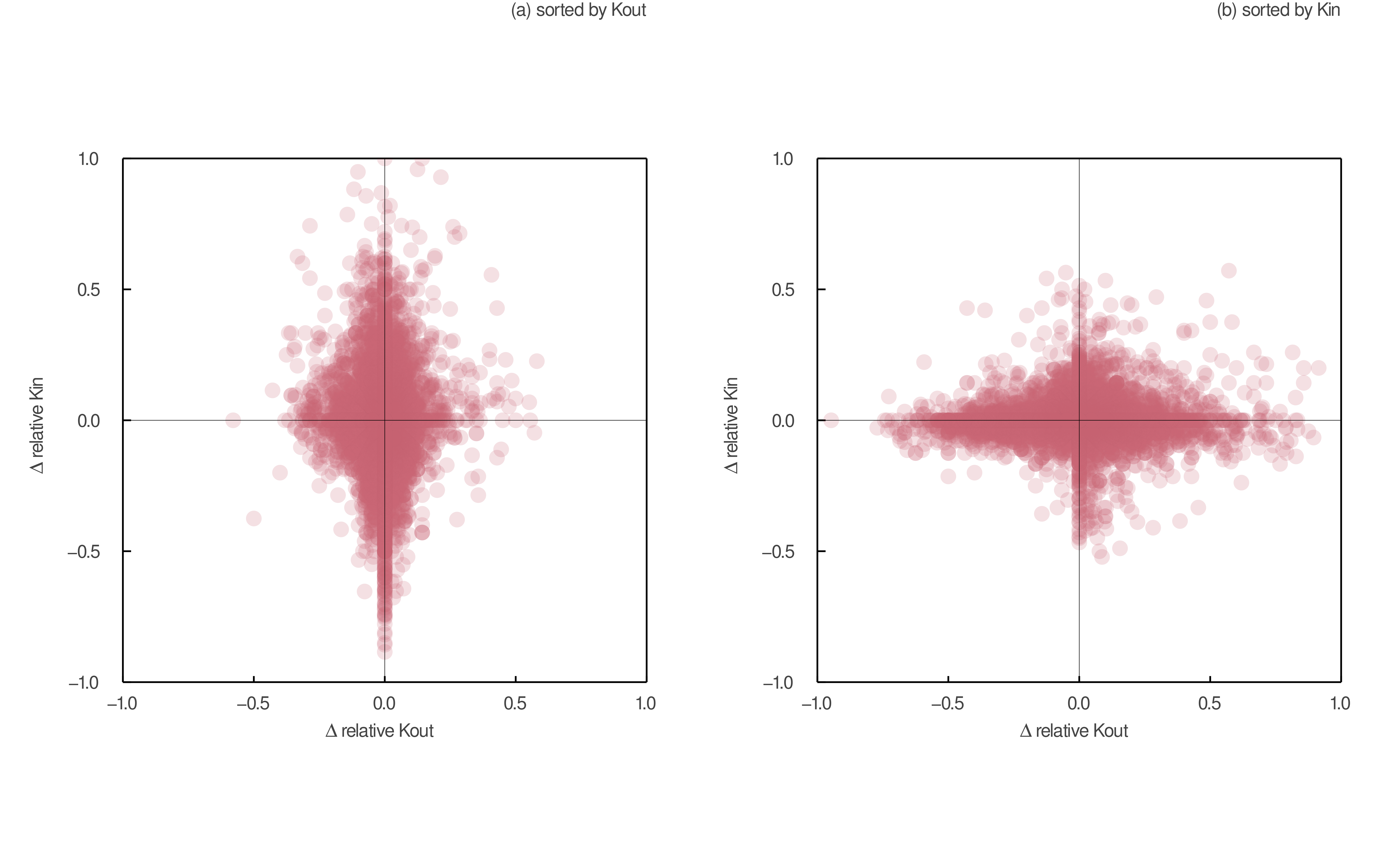}
\caption{Difference between predicted and empirical values for the
relative number of predators \(k_{in}\) and the relative number of prey
\(k_{out}\) when species are ordered according to (a) their out-degree
and (b) their in-degree. Empirical networks include most food webs
archived on Mangal, as well as the New Zealand and Tuesday lake datasets
(our complete dataset). The predicted joint degree sequences were
obtained after sampling one realization of the joint degree distribution
of maximum entropy for each network, while keeping the total number of
interactions constant. In each panel, each dot corresponds to a single
species in one of the networks.}\label{fig:kin_kout_diff}
}
\end{figure}

\begin{figure}
\hypertarget{fig:heatmap}{%
\centering
\includegraphics{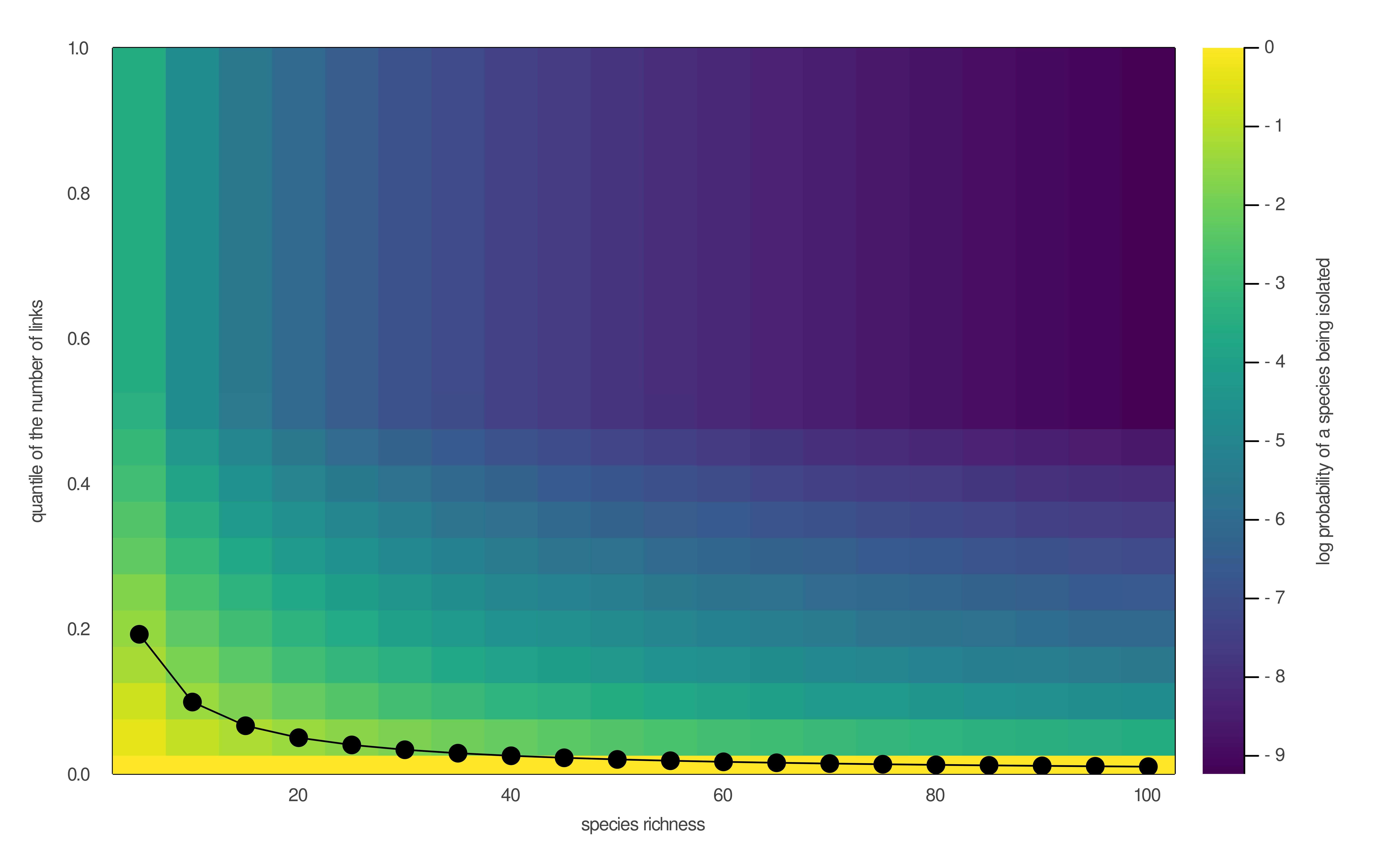}
\caption{Probability that a species is isolated in its food web
according to the degree distribution of maximum entropy. We derived many
degree distributions of maximum entropy given a range of values of \(S\)
and \(L\), and plotted the probability that a species has a degree \(k\)
of \(0\) (log-scale color bar). Here species richness varies between
\(5\) and \(100\) species, by increment of \(5\) species. For each level
of species richness, the numbers of interactions correspond to all
20-quantiles of the interval between \(0\) and \(S^2\). The black line
marks the \(S-1\) minimum numbers of interactions required to have no
isolated species.}\label{fig:heatmap}
}
\end{figure}

\begin{figure}
\hypertarget{fig:entropy_dist}{%
\centering
\includegraphics{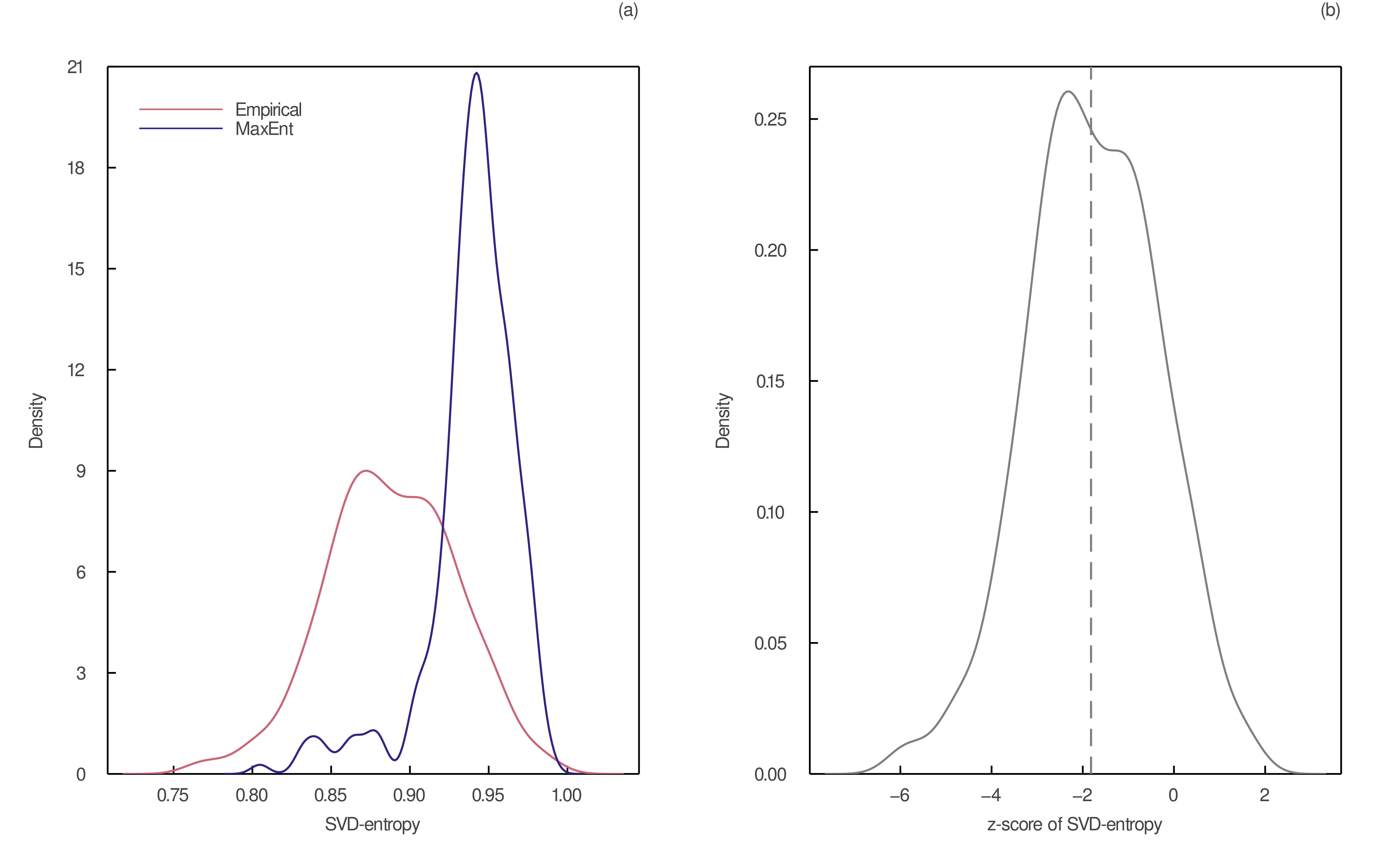}
\caption{(a) Distribution of the SVD entropy of empirical and maximum
entropy food webs. Empirical networks include most food webs archived on
Mangal, as well as the New Zealand and Tuesday Lake food webs (our
complete dataset). Maximum entropy networks were obtained using the type
II heuristic MaxEnt model based on the joint degree sequence. (b)
Distribution of z-scores of the SVD entropy of all empirical food webs.
Z-scores were computed using the mean and standard deviation of the
distribution of SVD entropy of MaxEnt food webs (type II heuristic
MaxEnt model). The dash line corresponds to the median
z-score.}\label{fig:entropy_dist}
}
\end{figure}

\begin{figure}
\hypertarget{fig:entropy_size}{%
\centering
\includegraphics{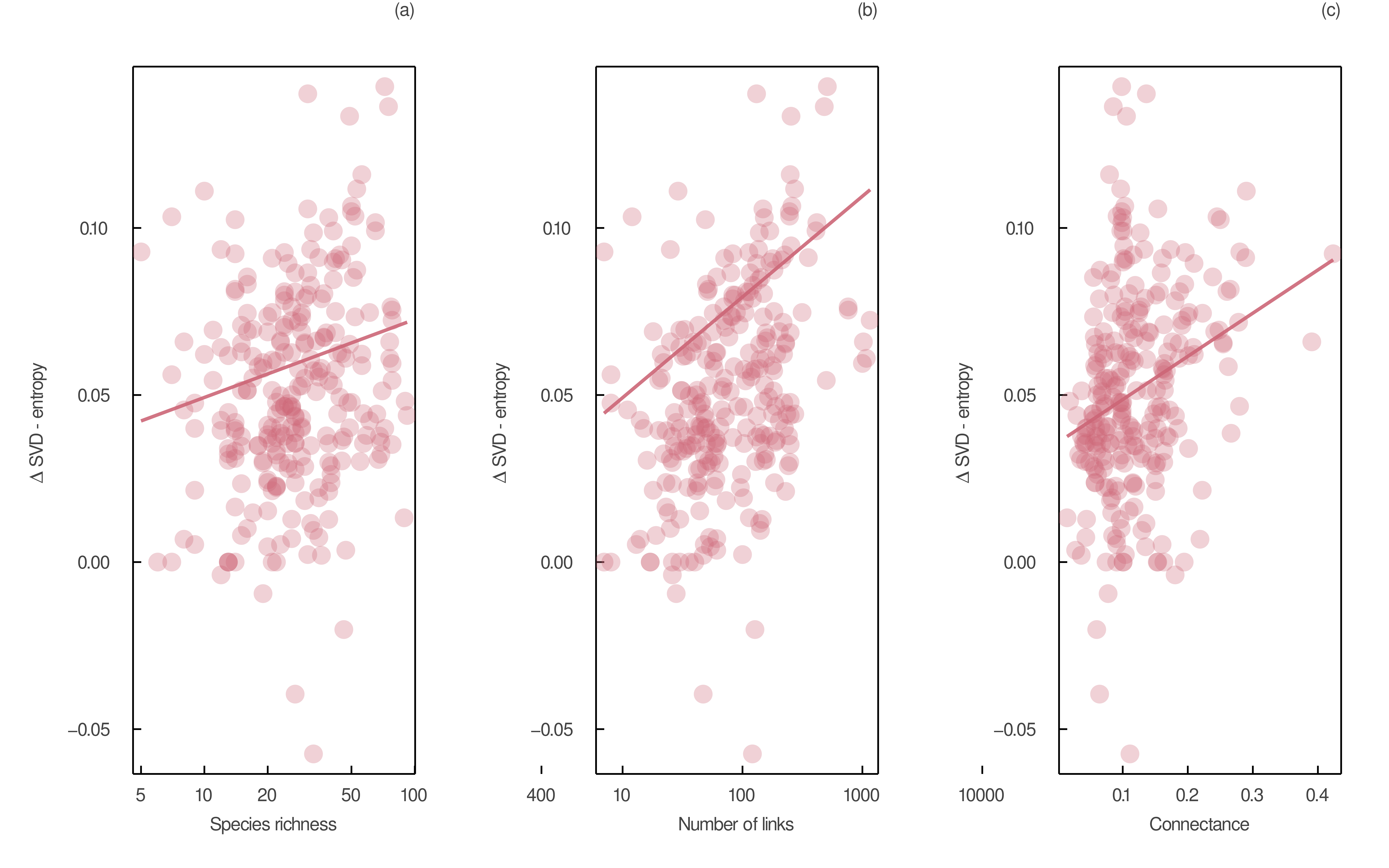}
\caption{Difference in SVD entropy between maximum entropy and empirical
food webs as a function of (a) species richness, (b) the number of
interactions, and (c) connectance. Empirical networks include most food
webs archived on Mangal, as well as the New Zealand and Tuesday Lake
food webs (our complete dataset). Maximum entropy networks were obtained
using the type II heuristic MaxEnt model based on the joint degree
sequence. Regression lines are plotted in each
panel.}\label{fig:entropy_size}
}
\end{figure}

\begin{figure}
\hypertarget{fig:measures_richness}{%
\centering
\includegraphics{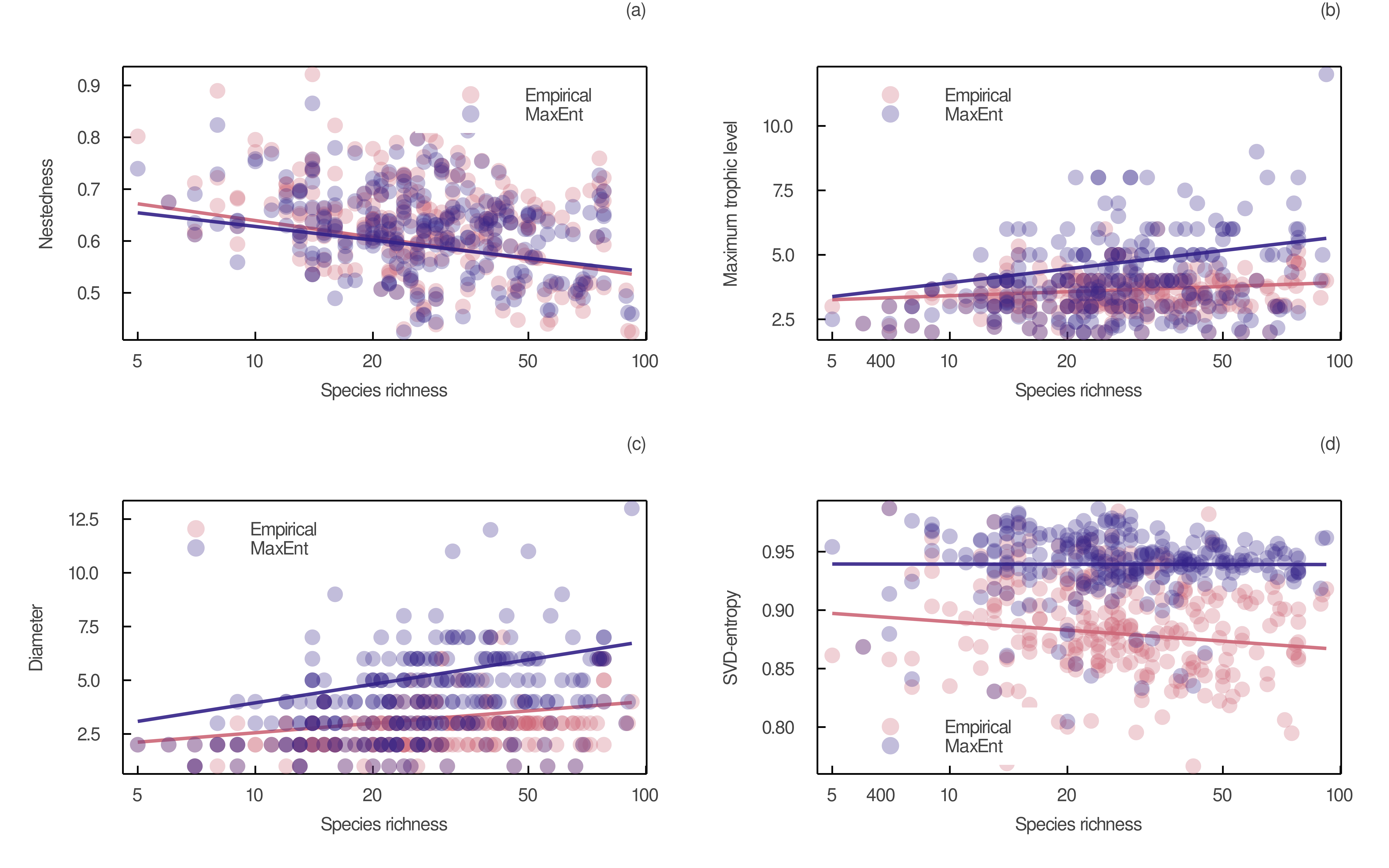}
\caption{Structure of empirical and maximum entropy food webs as a
function of species richness. Empirical networks include most food webs
archived on Mangal, as well as the New Zealand and Tuesday Lake food
webs (our complete dataset). Maximum entropy networks were obtained
using the type II heuristic MaxEnt model based on the joint degree
sequence. (a) Nestedness (estimated using the spectral radius of the
adjacency matrix), (b) the maximum trophic level, (c) the network
diameter, and (d) the SVD entropy were measured on these empirical and
maximum entropy food webs and plotted against species richness.
Regression lines are plotted in each
panel.}\label{fig:measures_richness}
}
\end{figure}